Manuscript for arXiv 2026-10-16

# Interception Focusing of Colloid Transport in Granular Porous Media

William P. Johnson[1,*] Noah G. Willis[1], Luis A. Ullauri[1], Bashar M. Al-Zghoul[2], Diogo Bolster[3]

[1]University of Utah, Geology and Geophysics, Salt Lake City, Utah 84112, USA

[2] University of California Davis, Civil and Environ. Eng., Davis, California 95616, USA

[3]University of Notre Dame, Civil and Environmental Engineering and Earth Sciences, Notre Dame, Indiana 46556, United States

**Abstract**

Colloid transport and retention in environmental granular porous media underlie groundwater resource protection from, and remediation using, biological and non-biological nano- and micro-particles (colloids). During transport, colloids periodically intercept grain surfaces and translate along them at separation distances across several tens of nm corresponding to the secondary minimum between the opposing forces of longer-range of van der Waals (VDW) attraction and shorter-range electric double layer (EDL) repulsion. Relatively weak diffusive translation of colloids supports their continued translation in secondary minimum association until they are expelled to bulk fluid at rear flow stagnation zones on grain leeward sides. We herein show that this dynamic focuses colloids into a network of fluid streamlines (incomplete mixing lamellae) that enhances their subsequent likelihood of interception with downgradient grains, i.e., increases the so-called collector efficiency ($\eta$). Interception focusing of colloids in granular porous media explains well-observed outcomes of colloid transport under environmental conditions, including nonmonotonic nonexponential profiles of retained colloids and the non-Markovian nature of extended tailing of low colloid concentrations following initial elution. Interception focusing is a fundamental process that distinguishes colloids from solutes and renders inapplicable to colloids the assumption that is often invoked for solutes of complete pore scale mixing in granular porous media.

## Introduction

*Impacts of Size and Topology*

Colloid and solute size ranges break at approximately several nm, and this distinction drives greater diffusive translations for solutes and longer-reaching surface interactions for colloids, which play out in their attachment to surfaces, e.g., often reversible for solutes and largely irreversible for colloids in the absence of perturbations in flow and solution chemistry (e.g., Pazmino et al., 2014; Rasmuson et al., 2019; VanNess et al., 2019; Johnson and Pazmino, 2023).[1-4] It also manifests in their interaction with flow field topology, i.e., zones of flow divergence and convergence at forward and rear flow stagnation points (FFSP and RFSP), respectively on sediment grains (Figure 1). FFSP have long been recognized as primary zones of colloid delivery to the near surface fluid domain as enveloped in classic colloid filtration theory (CFT) (Rajagopalan and Tien, 1976).[5] RFSP have more recently been recognized as zones of colloid expulsion to bulk fluid from the near surface fluid domain (e.g., Johnson et al., 2018),[6] where the near-surface fluid domain corresponds to colloid-surface separation distances sufficiently small (e.g., < 200 nm) to allow the farthest-reaching interaction (van der Waals) to bring the colloid trajectory toward the grain surface.

RFSP and FFSP are apparent in pore scale colloid transport experiments conducted in 2D flow cells (Figure 1) (Johnson et al., 2025)[7] as delineated by colloid locations (bottom panel) and solute (fluorescein) fluorescence intensities (top panel). Dark lamellar zones with fluorescein tracer concentrations below optical detection (Figure 1, top panel) emanate from RFSP in the left-to-right flow field, wherein grains (cylinders) appear as black circles and bright green zones correspond to high fluorescein concentrations, consistent with preceding experiments describing such incomplete mixing lamellae (IML) (e.g., de Anna et al., 2013; LeBorgne et al., 2013).[8,9] IML may merge downstream and may also terminate at FFSZ of down-gradient grains to form grain-connecting lamellae (GCL) that may hold significance for colloid delivery to grains (e.g., Johnson, 2020; Johnson et al., 2025).[7,10] IML as evidenced by solutes are transient, on the order of several seconds in these experiments, since solute diffusion laterally across streamlines drives complete pore scale mixing of solute.

In contrast to solutes, colloids experience IML persistently due to the multiple orders-of-magnitude lesser diffusive translation of colloids relative to solutes (Figure 1, bottom panel). Fluid streamlines, FFSP, RFSP, and IML are equivalently by solute and colloid trajectories (Figure 1, top versus bottom panel) despite the factor-of-400 flow rate difference utilized for these experiments concerning colloid and solute transport. The non-transient impact of IML on colloid transport in granular media (Figure 1) renders the often-invoked assumption of complete pore scale mixing or solutes fully inapplicable to colloids, and concurrently impacts the subsequent trajectories of colloids that intercept the near surface fluid domain but fail to attach, as described below.

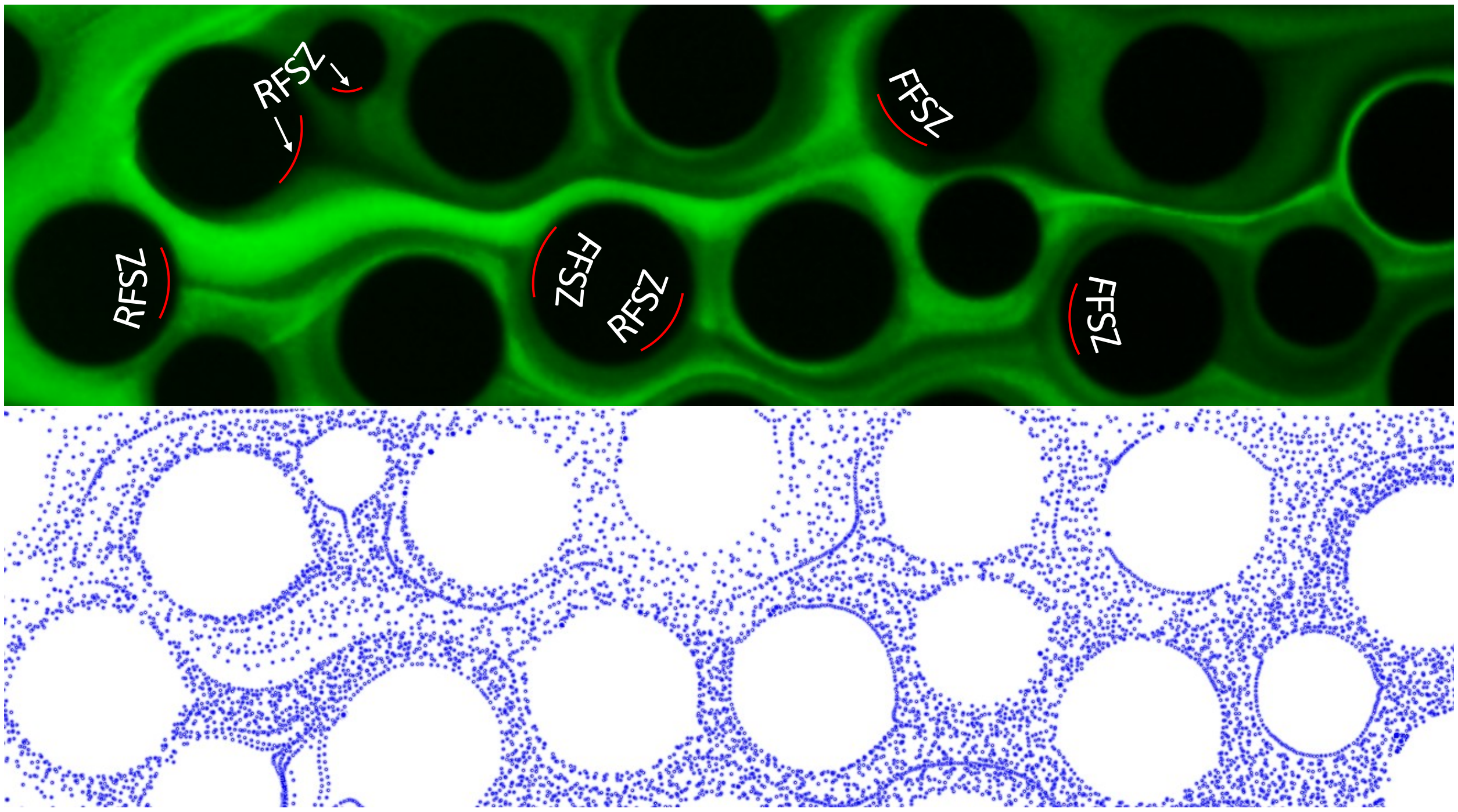

**Figure 1**. Images of portion of 2D silica porous media with channel thickness of 200 um with cylinders (black in top panel, white in bottom panel) of 100, 150, and 200 μm diameters and porosity 52.7%. Top panel: Injection of fluorescein tracer (40 ppm, IS: 5.4 mM, pH: 6.5) flowing from left to right at 8000 μL/min, 5× magnification. Tracer brightness is proportional to concentration, revealing incomplete mixing lamellae (IML) that emanate from the rear flow stagnation zone (RFSZ) of grains (cylinders). IML emanating from an upstream collector may connect to the forward flow stagnation zone (FFSZ) of a downstream collector, forming grain connecting lamellae (GCL). Bottom panel: Colloid locations (blue dots) (1 μm diameter

carboxylate-modified polystyrene latex) delineating trajectories moving from left to right with solution flow at 15 µL/min, ionic strength at 5.4 mM, pH at 6.5, 5× magnification.

*Impacts of Unfavorable Attachment Conditions*

Colloid delivery to the near surface fluid domain via FFSP, and expulsion back to bulk fluid via RFSP, is most apparent under so-called unfavorable attachment conditions in which the colloid and grain surfaces are like-charged (typically both negative) thereby producing electric double layer (EDL) repulsion that decays exponentially with increased separation distance (Israelachvilli, 2011).[11] Because VDW attraction decays according to a power law relationship with increased separation distance, VDW attraction predominates over EDL repulsion at the closest separation distances (e.g., < 1 nm) and at large separation distances (e.g., greater than a few tens of nm) (Israelachvilli, 2011).[11] Superposition of EDL and VDW interactions under unfavorable conditions produces a colloid-surface interaction profile that is strongly attractive at separation distances < ~ 1 nm (primary minimum), repulsive at intermediate separation distance between one and several tens of nm (repulsive barrier), and weakly attractive beyond several tens of nm (secondary minimum) out to the limit of VDW attraction (~ 200 nm). Interception of a grain surface is defined by colloid entry into the near surface fluid domain where VDW (the longest-range interactions) become significant, e.g., separation distances of < 200 nm (Johnson and Pazmino, 2023).[4]

Unfavorable conditions are characteristic of environmental colloid transport (Johnson and Pazmino, 2023).[4] The process of colloid attachment following interception therefore involves residence "in" the secondary minimum during translation across grain surfaces at colloid-grain separation distances typically around a few tens of nm. Local elimination of the repulsive barrier via the existence of nanoscale zones of attractive charge on the surfaces termed heterodomains (Johnson and Pazmino, 2023)[4] allows colloid translation toward the primary minimum where attachment can occur. The mechanics of colloid interaction with heterodomains, the potential physicochemical origins of heterodomains, and the torque balance for attachment are described elsewhere (e.g., Johnson and Pazmino, 2023).[4] However, the upshot is that along their near-surface trajectories colloids experience a range of repulsive barrier sizes between maximum and absent corresponding to the degree of overlap of the zone of colloid-surface interaction (ZOI) with heterodomains (Johnson and Pazmino, 2023).[4] Colloid transport and retention in granular media under environmental or other unfavorable attachment conditions is therefore characterized by delivery of a fraction ($\eta$) of the approaching population to the near surface fluid proximal to the grain FFSP, slow translation of near surface colloids along grain surfaces, attachment of a fraction ($\alpha$) of the near surface colloid population, and expulsion of non-attaching near surface colloids back into bulk fluid proximal to grain RFSP (e.g., Johnson et al., 2025).[7]

*Impacts of Interception History*

The above dynamic involving repeated interception followed by attachment or expulsion under unfavorable conditions is reflected in classic CFT, but only under limiting conditions, which are that: a) the value of $\alpha$ approaches unity (favorable conditions); or b) the values of $\alpha$ and $\eta$ are unchanged as colloids progress to subsequent interceptions after failing to attach, as described in the interception history paradigm (Volponi et al., 2025; Johnson et al., 2025; Al-Zghoul et al., 2025).[7,12,13] Under condition (a), favorable conditions wherein $\alpha\approx1$, the vast majority of colloids attach upon their first interception and thereby produce exponential RP that are well-described by CFT (Johnson and Pazmino, 2023).[4] Under condition (b), unfavorable condition with constant $\alpha$ and $\eta$, only a fraction ($\alpha$) of the colloid population attaches upon first interception to form an exponential RP, whereas the remainder of the population attaches after multiple interceptions to form gamma-shaped RPs that together with the single-intercepting attacher RPs form an overall exponential RP with log-linear slope governed by $\alpha$ and $\eta$ (Al-Zghoul et al., 2025; Volponi et al., 2025; Johnson et al., 2025) (Figure SI-1):[7,12,13]

$$\ln\left(\frac{N^{att}(x)}{N_0}\right)=\ln(\alpha)-\left(\frac{\alpha k_f}{v}\right)*x \qquad (1)$$

where $N_0$ and $N^{att}(x)$ are the number of colloids injected at source and the number attached at a given downgradient distance, $v$ is the average pore water velocity, $x$ is the transport distance from the colloid source, and where the interception rate ($k_f$) in a 2D medium is:

$$k_f=-\frac{4(1-\theta)^{\frac{1}{2}}}{\pi d_c}vLn(1-\eta) \qquad (2)$$

where $\theta$ is the pore network porosity, and $d_c$ is diameter of the cylindrical grains, as derived in the Supporting Information. However, experimentally observed RPs under unfavorable conditions are well known to take nonexponential shapes that, according to the interception history paradigm, may be multiexponential or nonmonotonic forms depending on whether $\alpha$ (or $\eta$) decreased or increased, respectively, with interception order (Figure SI-2) (Volponi et al., 2025; Johnson et al., 2025, Al-Zghoul et al., 2025).[7,12,13]

The objective of this paper is to demonstrate that colloid interception focuses colloid expulsion proximal to RFSP and thereby directs colloids into an IML network that emanates from RFSP. We observed that intercepting colloids are focused proximal to RFSP prior to expulsion back to bulk fluid thereby aligning them with ICL, and we observed that the net distance traveled between interceptions decreased by approximately a factor of two for subsequent relative to first interceptions. We observed an approximate factor-of-two increase in $\eta$ for collectors downstream versus upstream of GCL, and we documented the prevalence of GCL among IML. We demonstrate that a factor-of-two increase in $\eta$ is sufficient to produce nonmonotonic RP consistent with that observed experimentally (Johnson et al., 2025).[7] We discuss the relationship of repeated interception to extended tailing in breakthrough-elution concentration histories (BTEC) and the relationship of interception focusing to the non-Markovian nature of extended tailing. We speculate that multiexponential RPs potentially arise from a lesser fraction of GCL among IML as dictated by grain packing and porosity and/or a counteracting decrease in $\alpha$ with interception order. We assert more broadly that interception focusing into IML and particularly GCL is a fundamental process that distinguishes colloid transport from solute transport in granular porous media.

**Methods**

*Porous Media and flow cell*

Transport experiments were performed in a two-dimensional porous media composed of silicon cylinders (200 µm length) with varied diameters (50, 100, and 200 µm) placed randomly to produce 37% porosity (Figure SI-3). All surfaces were coated with $SiO_2$ (100 nm thickness). The rationale for a 2-D porous media was that colloid trajectories are visible along their entire length in a single focal plane across a relatively large area. The 2-D porous media was fabricated at the University of Utah Nanofabrication Laboratory as described in Johnson et al., (2025).[7] Cylinder radii decreased 1 µm for every 20 µm in depth, yielding a 10 µm smaller diameter at the cylinder bottoms relative to their tops, or a 20%, 10%, and 5% decrease for the 50 µm, 100 µm, and 200 µm diameter cylinders, respectively, as an unavoidable outcome of the fabrication process. A 100 nm-thick layer of $SiO_2$ was deposited over the patterned cylinders via plasma-enhanced chemical vapor deposition (Oxford 80+ PECVD system) at 300°C. The wafers were diced into individual chips (Disco DAD 641 – dicing saw).

The porous medium and flow cell were cleaned for each transport experiment via the standard cleaning (SC-1) procedure (Kern and Puotinen, 1970),[14] wherein the flow cell and cylinder array were submerged in a Milli – Q, $H_2O_2$ (30% w/w), and NH4OH (28% w/w solution) in a volume ratio of 5:1:1 at 75 – 80 °C for 10-15 minutes.

*Flow system and flow conditions*

Flow was controlled using a syringe pump (PHD 2000 Programmable, Harvard Apparatus). Pore water velocity was set to maximize the number of frames defining the trajectories while supporting sufficient colloid flux to obtain target numbers of attached colloids. Average pore water velocities 19 and 9.7 m/day were used for experiments performed under favorable and unfavorable conditions, respectively. The factor-of-two difference in the flow rates was not intentional, however, the impact of this velocity difference on $\eta$ is negligible relative to the impact of favorable versus unfavorable conditions on $\alpha$ (e.g., Johnson and Pazmino, 2023).[4]

Injection was terminated when a target number of colloids (e.g., approximately 200) were attached within the field of view. The elapsed time when the injection solution was switched from colloid to colloid-free solution was recorded, and the count of attachments included any that attached after switching the solution. The constancy of the flow field (colloid trajectories) was checked across one-minute intervals every five minutes across the experiment duration (30 to 90 minutes). In the event of a flow field shift (due to air bubble entry to the system), data acquisition and analysis was terminated.

*Colloidal suspensions*

Experiments employed carboxylate-modified polystyrene latex (AML and CML) fluorescent ($\lambda_{ex}$ = 505, $\lambda_{em}$ = 515 nm) 0.95 µm diameter microspheres (MS) (FluoSpheres™, Thermo Fisher

Scientific Inc., Eugene, OR) to produce unfavorable interactions with the flow cell silica surfaces. The injected concentration ($C_0$) was $2.6 \times 10^4 \pm 20\%$ MS/mL, a relatively low concentration that allowed distinction of individual colloid trajectories in the field of view while allowing sufficient attachment. The microsphere suspension concentration was determined via vacuum filtration of colloid solution as described in Johnson et al. (2025).[7]

The CML suspension had an ionic strength (IS) of 6.0 mM, as adjusted by NaCl addition in 2.2 mM MOPS (3-morpholinopropane-1-sulfonic acid) buffer (Sigma-Aldrich Corp.) with solution pH set to 6.5 using 0.5 M NaOH.

CML electrophoretic mobilities (EPM) were measured in suspensions using a ζ-potential analyzer (Mobiuζ, Wyatt Technology Corp., Santa Barbara, CA), and ζ-potentials were calculated from EPM via the Smoluchowski equation. Measured CML and AML ζ-potentials under unfavorable and favorable conditions were -64.5 ± 3.4 and 12.1 ± 0.5 mV, respectively. Silica ζ-potentials were assumed to be -70 mV under both unfavorable and favorable conditions for simplicity (Trauscht et al., 2015).[15]

*Image acquisition*

Colloid transport in the porous medium was tracked by using fluorescence microscopy (Leica DMi8 inverted microscope, Leica Microsystems, Wetzlar, Germany) with a 2.5x objective (HC FL PLAN 2.5x/0.07 DRY, Leica Microsystems, Wetzlar, Germany) and LED5 fluorescence light source (Leica Microsystems, Wetzlar, Germany), a FITC bandpass filter for excitation (475 nm), and a (527 nm) bandpass filter for emission (Leica Microsystems, Wetzlar, Germany).

Magnification (2.5x) was optimized to maximize the field of view while also maximizing trajectory resolution. The numerical aperture of the objective was 0.07, and the medium refractive index was 1. Images and video recordings were acquired via a K8 Scientific CMOS Camera (Leica Microsystems, Wetzlar, Germany) and analyzed using the image analysis software Leica Application Suite X (LAS X) (Leica Microsystems, Wetzlar, Germany). Images were acquired every 0.064 s (~15 fps) with a digitalization of 12-bits and a sensor format of 2048 px x 2048 px (4.2 Megapixel).

Colloid trajectories were tracked at the flow cell inlet in a square field of view (5362 µm per side) (Figure 2) with a focal depth of 108 µm. The focal plane was positioned at the mid-depth of the porous medium, approximately 100 µm away from the top and bottom of the cylinders. Focal depth was approximately 200 µm, which was the entire depth of the flow cell. Colloids deposited on bounding chamber surfaces (> 7 µm separation distance from cylinders) were not considered for analysis. These represented less than 12% of attachments under both favorable and unfavorable conditions. Because of the large number of attachments across the field of view under favorable conditions, only the upper one fifth of the field of view was used in the favorable condition analysis.

*Colloid Trajectory Analysis*

For each attached colloid, the image stack was played backwards frame by frame to identify the entry location into the field of view. This allowed cropping the overall video to a subset containing the full colloid trajectory from entry to attachment. The x-y coordinates and frame number for each acquired image was obtained using LAS X software (LEICA, Inc.). A bright field optical image of the porous media was used to determine coordinates for the cylinder centers. Circles with predefined radii (25, 50, and 100 µm) were drawn around the cylinders (AutoCAD) on the brightfield image. Center x-y values (AutoCAD Data Extraction tool) between the brightfield and fluorescence images deviated modestly (between 5 and 10 µm) due to slight differences in cell inclination between brightfield and fluorescence images. The 5 to 10 µm deviation was larger than the pixel size at 2.5x magnification (2.6 µm), therefore cylinder center x-y values in the fluorescence images were updated to match the brightfield image to account for distortion from slight flow cell inclination.

*Interception identification*

A combination of frame-by-frame colloid velocities and apparent colloid-cylinder separation distances were used to identify colloid interception of cylinders in experiments. The minimum recordable colloid velocity in experiments was 40 µm/s corresponding to the pixel size (2.6 µm) divided by the acquisition time (0.064 s). The colloid velocity values were normalized by their maximum values to identify portions of the trajectory with low velocity as an indicator of interception, as described in Johnson et al. (2025).[7] Apparent colloid-cylinder separation distances were used secondarily to velocity to identify interceptions because the optical resolution (2.6 μm) was too coarse to allow identification of colloid entry to the near surface fluid domain where colloid-surface interactions become significant, corresponding to separation distances less than 200 nm (Johnson, 2020).[10] Apparent colloid-cylinder separation distances were obtained using trajectory x-y values and corrected cylinder centers, accounting for cylinder and colloid radii.

Colloid entry to the near-surface fluid domain (interception) was defined as colloid velocities < 2% of the average colloid velocity across its trajectory and colloid-cylinder separation distances < 7 μm for 7 out of 10 consecutive translations. This criterion for interception is the same as that used by Johnson et al. (2025),[10] except six duplicate trajectories were removed from the set of 204, which produced a qualitatively equivalent RP (Figure SI-4). Colloid exit from the near-surface fluid domain was defined by the same criterion except that the threshold colloid-cylinder separation distance was 5 μm rather than 7 μm. Near-surface trajectories corresponding to the entry and exit criteria were plotted and checked visually for reasonable entry and exit locations.

*Solute and Colloid Comparison*

Fluorescein transport experiments for comparison with colloid transport experiments were performed in a flow cell wherein cylindrical collectors of 100, 150, and 200 μm diameters were placed to produce 52.7% porosity (Figure 1). All other flow cell components were as described above for the 37% porosity flow cell. Injection of fluorescein (40 ppm, IS: 5.4 mM, pH: 6.5) at a flow rate of 8000 μL/min was recorded by LAS-X software at 5× magnification (field of view: 2681 μm × 2681 μm). Injection of CML (5.4 mM NaCl, pH: 6.5, $2.6\times10^4$ microspheres/mL) at 15 μL/min was recorded by the same method, magnification, and field of view as for fluorescein.

*Mechanistic Colloid Trajectory Simulations*

Mechanistic pore-scale colloid trajectory simulations in the 2D flow cell were performed as described in Johnson et al. (2025) and Al-Zghoul et al. (2024).[7,16] Normalized colloid velocity and separation distance thresholds were tested as indicators of colloid interception of the near surface fluid domain as described in Johnson et al. (2025).[7] Mechanistic colloid trajectory simulations to examine the impact of interception location on $\alpha$ were performed in a Happel sphere-in-cell collector as described in detail in Johnson and Pazmino (2023).[4]

**Results and Discussion**

*Interception focusing in 2D continuum-scale experiments*

For the digitized trajectories of the 198 attached colloids in the 2D experiments (Johnson et al., 2025),[7] the net distance travelled between successive interceptions was larger prior to the first interception relative to subsequent interceptions (Figure 2), indicating that interception focusing during the first interception reduced the number of collectors passed prior to subsequent interceptions, and suggesting that interception focusing impacted $\eta$.

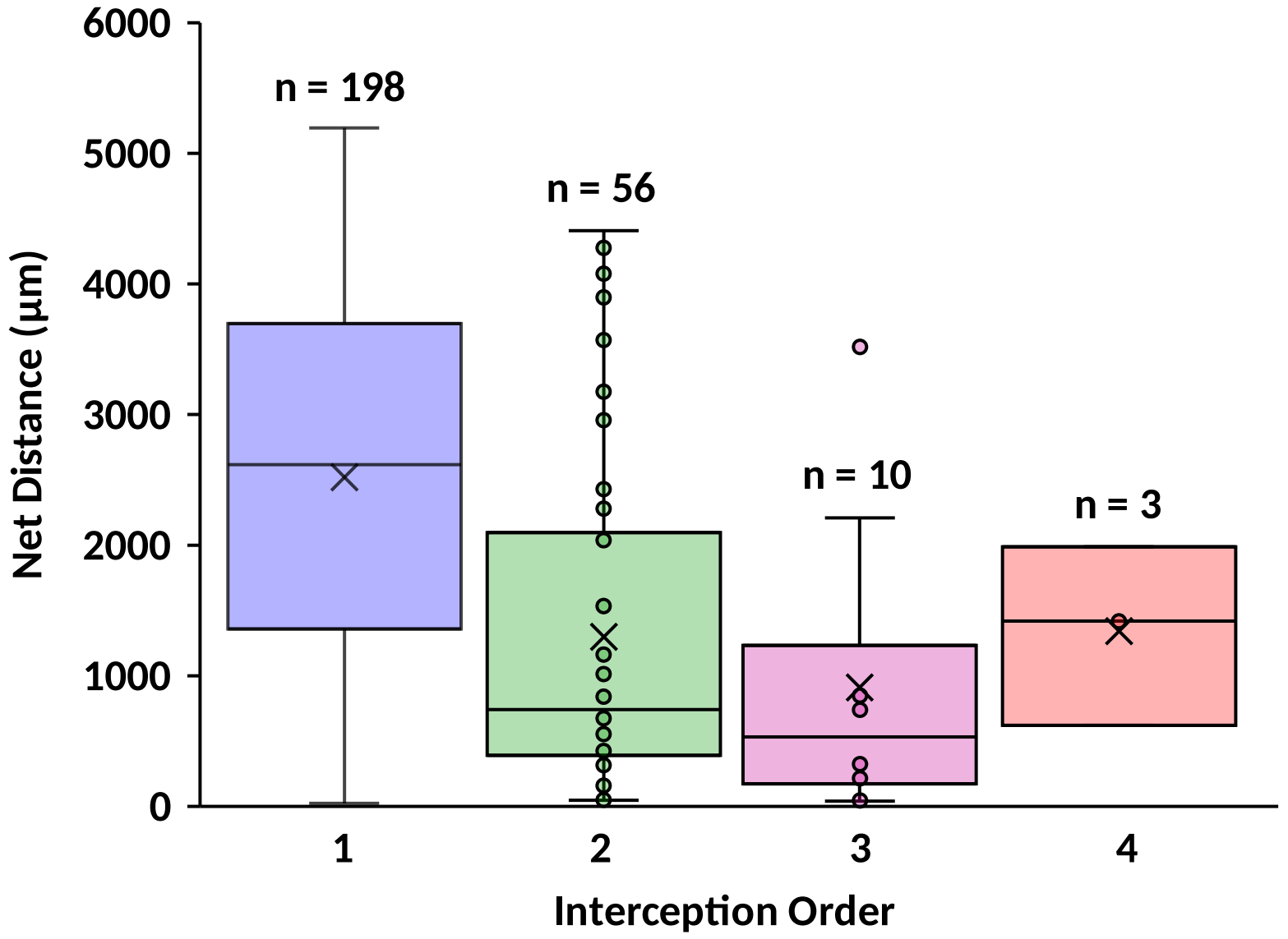

**Figure 2**. Net distance travelled prior to each interception of the near surface fluid domain for the set of 198 attaching colloids. Error bars reflect standard deviation in the net distance traveled each for the n individuals that underwent the corresponding order of interception.

Colloid interception locations relative to FFSP were examined in terms of angles between the respective poles to the grain center from the interception and FFSP locations (F-I angles) (Figure 3). F-I angle mean and standard deviation are shown as a circle and arc, respectively, on the perimeter of an example grain (cylinder) for single-intercepting attachers (blue) and multiple-intercepting attachers (red) (Figure 3, panel tops) and in box plots (Figure 3, panel bottoms). Differences between F-I angles for first relative to subsequent interceptions were statistically insignificant in two-tailed Mann-Whitney U tests.

A mechanism for interception focusing is suggested by angles between the location of near surface exit and the RFSP (E-R angles) (Figure 3). Under unfavorable conditions in experiments,

F-I variance was roughly twice that of E-R variance (factor of 2.1), a difference confirmed by F-test, Brown-Forsythe, and Fligner-Killeen tests (all $p < 0.01$). The narrower range of E-R relative to F-I angles is observed also under favorable conditions in experiments, albeit with a far lesser number of subsequent interceptions (Figure 3). We submit that expulsion of colloids proximal to RFSP placed them onto the IML network, in which a portion of the IML were GCL that, we posit, enhanced the subsequent likelihood of colloid interception, as examined below.

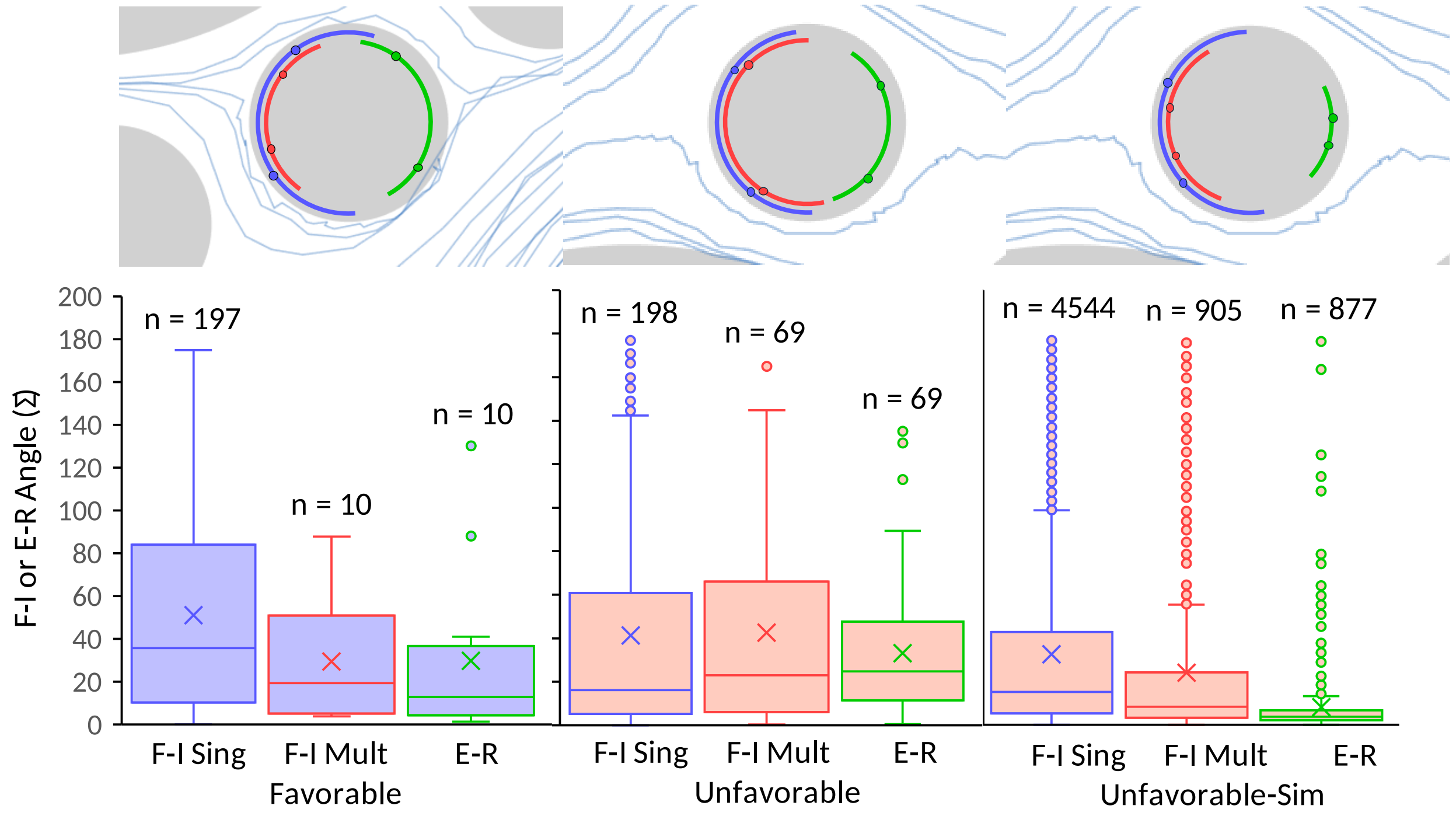

**Figure 3**. FFSP-to-interception (F-I) angles for first and subsequent interceptions for favorable conditions (panel a), unfavorable conditions (panel b), and simulated under unfavorable conditions (panel c). Panel tops: Superimposed example colloid trajectories (blue lines) under favorable (left) and unfavorable (right) conditions in a representative portion of the 2D flow cell. Circles and arcs on grain (cylinder) peripheries indicate mean and standard deviation, respectively, of F-I (FFSP to interception) angles for first (blue) and subsequent (red) interceptions, shown on each side of the FFSP. E-R (exit to RFSP) angle mean and standard

deviation are shown (green) on either side of the RFSP. Panel bottoms: Boxplots of F-I angles for intercepted colloids under favorable (blue fill) and unfavorable (red fill) conditions. Within each condition, the left box represents first interceptions (blue outline), the middle box represents subsequent interceptions (red outline), and the right box represent E-R angles (green outline). Elements of the box and whisker plots include the median (horizontal line inside the box), mean ("x" inside the box), the 25% and 75% quartiles (box bottom and top, respectively), minimum and maximum data points excluding outliers (whisker ends).

Angles corresponding to the distance traveled between interception and attachment (I-A angles) were larger under unfavorable relative to favorable conditions for both single- and multiple-intercepting attachers (Figure SI-5), reflecting the longer path of translation prior to attachment across the grain near surface under unfavorable conditions.

To determine whether interception focusing altered the collector efficiency ($\eta$) requires determining which colloid trajectories entered (# approached) the fluid shell associated with any given grain (cylinder in 2D) and which entering colloids intercepted the near-surface fluid domain (# intercepted) wherein $\eta$ = #intercept/#approach. Fluid envelopes (shells) surrounding grains (cylinders) in the 2D flow cell were defined to honor porosity, where in CFT in granular media, the ratio of the fluid shell radius ($R_b$) to grain radius ($R_g$) is equal to the cube root of the complement of the porosity ($\theta$) such that $R_b/R_g = (1-\theta)^{1/3}$. For the 2D porous media composed of cylindrical collectors, the ratio of the fluid shell radius to grain radius that honors porosity is the square root of the complement of the flow cell porosity, i.e., $R_b/R_g = (1-\theta)^{1/2}$. However, the above formula generates large overlap of fluid envelopes in the 2D model due to the non-regular arrangement of multiple-sized cylinders (Figure SI-6, left panel). For this reason, all fluid shell radii were reduced by a factor of 4 to reduce overlap of adjacent fluid envelopes (Figure SI-6, right panel). Whereas this approach left a portion of the pore space unassociated with any given collector, it honored the ratio of fluid shell to grain radii dictated by the system porosity.

To further assess potential changes in $\eta$ from interception focusing in the 2D pore network, values of $\eta$ were determined for pairs of collectors linked by GCL to determine whether $\eta$ changed for collectors positioned upstream versus downstream of GCL. The criterion for interception in digitized trajectories was colloid velocity ≤ 2% of the average velocity of the colloid across its trajectory for 7 out of 10 consecutive translations for separation distances ≤ 7 μm (Johnson et al., 2025).[7] Interception for non-digitized (non-attaching) trajectories was done visually to identify approaching colloids (those that entered the fluid envelope) that qualitatively met the criterion employed by Johnson et al. (2025)[7] (video available at Zenodo via link provided in Data Availability Statement). Non-attaching colloid trajectories were examined visually because digitizing the non-attaching trajectories would have been intractably laborious and beyond the scope of our efforts since the number of non-attaching trajectories dwarfed the number of those that attached.

Six of seven randomly selected collector pairs connected by GCL showed values of $\eta$ that were higher for downstream relative to upstream collectors connected by GCL by a factor of 2.1 ± 1.2 (Figure 4, Table SI-1). The sole decrease observed among seven pairs examined was modest (factor of 0.7) and corresponded to the pair separated by a long distance (Figure 4, panel c, collectors 1 and 2), suggesting that interception focusing persists to a limited distance between collectors. That $\eta$ was observed to increase by an average factor of two from upstream to downstream collectors for six out of seven randomly selected pairs connected by GCL indicates that $\eta$ increased with interception order due to interception focusing. Notably, as observed more broadly (Figure 3), no upstream versus downstream trend was observed for F-I angles between the seven GCL-linked collectors (Table SI-5).

Given that only one or two attachments were observed per grain (cylinder), values of $\alpha$ carried errors on the order of a factor of two, which prevented evaluation of potential changes in $\alpha$ across the GCL. We reported previously (Johnson et al., 2025)[7] that the value of $\alpha$ increased with increasing interception order in the 2D flow cell experiments. However, we neglected in that determination of $\alpha$ the interceptions by non-attaching colloids, therefore the possibility that $\alpha$ changed with interception order in the experiments remains undetermined.

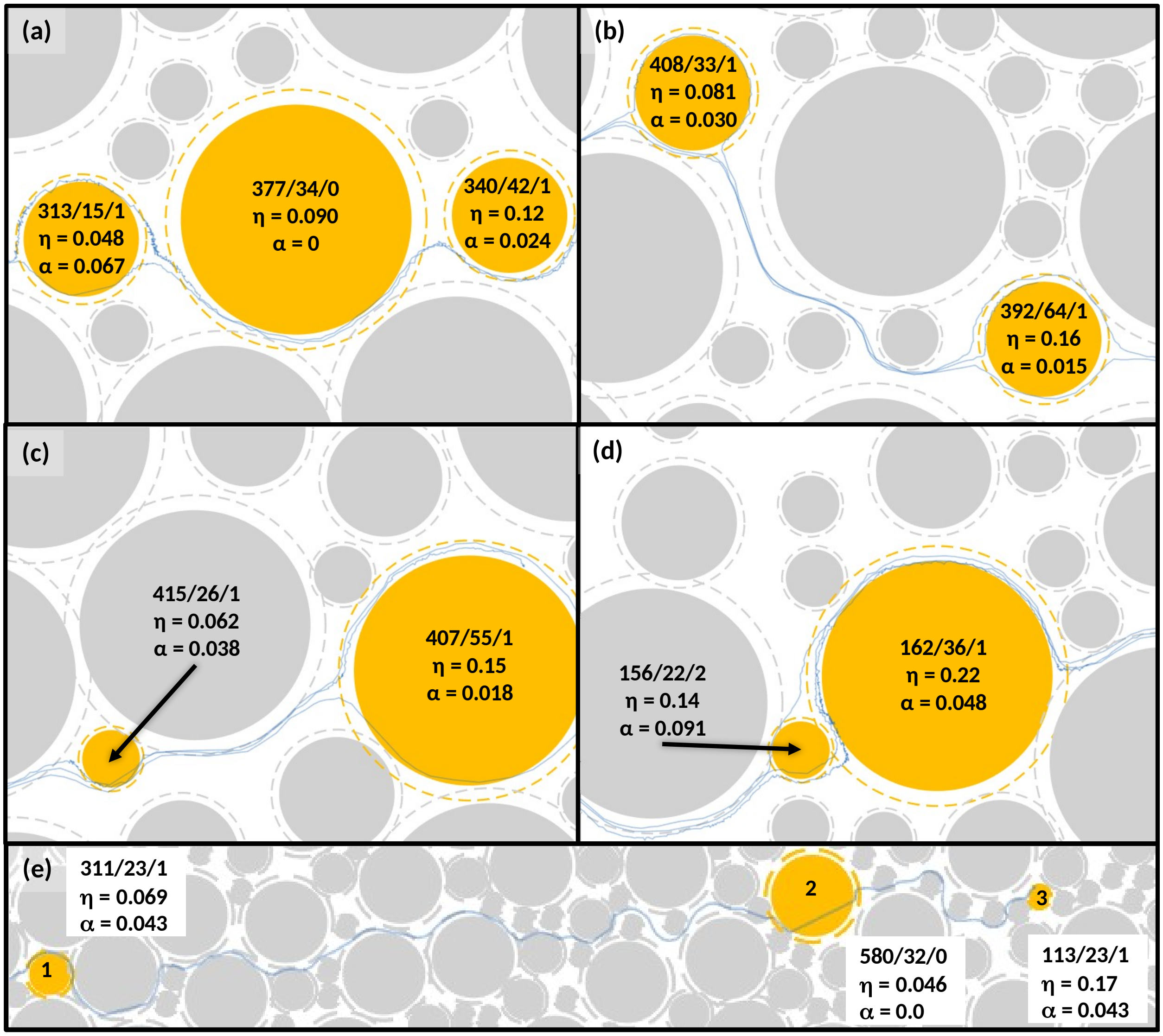

**Figure 4**. Select areas within the field of view of the 2D porous media showing grains connected by GCL (gold filled circles) and associated fluid envelopes (dashed gold circles) within which approaching and intercepting colloids are reported as approached/intercepted/attached, $\eta$, and $\alpha$ in the figure text. Occasional straight-line grain (cylinder) crossing reflected errors in digitized trajectories due colloid fluorescence dimming during interception (close proximity to cylinder). The endpoints of these cylinder crossings were checked in videos to confirm visually that they corresponded to a single colloid trajectory (Video available at Zenodo via a link provided in Data Availability Statement).

That $\eta$ was observed to increase by a factor of two from upstream to downstream for six out of seven randomly selected pairs of collectors connected by GCL prompted examination of the prevalence of GCL among IML in a portion (10%) of the experimental field of view (Figure 5). GCL were identified in experiment video recordings via pairs of collectors that were sequentially intercepted by colloids over distances within several grains, wherein interception was identified using the visual velocity criterion described above for determination of $\eta$. The prevalence of GCL among ICL was determined to be approximately half in the 10% field of view (Figure 5, left panel) as shown in detail for a subset area of 22 collectors (Figure 5, right panel). GCL appear to be common among ICL, supporting the possibility that focusing of colloid expulsion proximal to RFSP (Figure 3) focuses subsequent interception proximal to FFSP on downgradient grains.

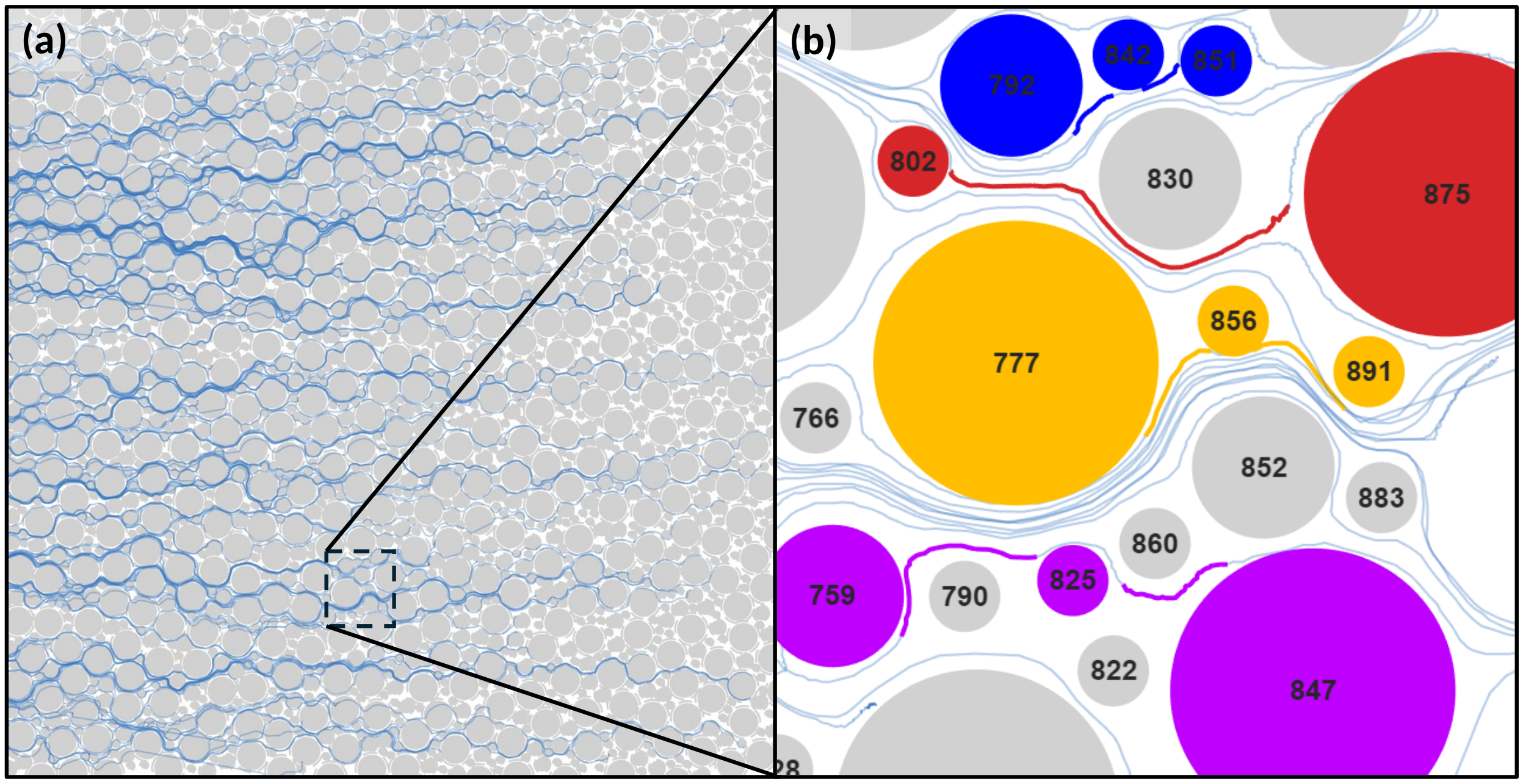

**Figure 5**. Panel (a): Experimentally observed trajectories of attached colloids (blue) overlayed on an image of cylindrical collectors (grey-filled circles) in the 2D pore network. Panel (b):

subset of 2D pore network showing pairs and triplets of collectors connected by GCL, as denoted by a common color among the pairs or triplets as well as portions of experimentally observed colloid trajectories that intercepted collector pairs/triplets.

*Interception focusing in Simulations*

Mechanistic colloid trajectory simulations in the 2D pore network (Al-Zghoul et al., 2024; Johnson et al., 2025)[16,7] suggested decreases in F-I angle for first relative to subsequent interceptions under the colloid velocity criteria for interception (7 out of 10 translations had velocities ≤ 2% of the average colloid velocity for separation distances ≤ 7 μm) (Figure 3, panel c). The mechanism of interception focusing in 2D simulations is also indicated to be expulsion proximal to RFSP as shown by the smaller E-R angles relative to F-I angles (Figure 3, panel c). In the 2D simulations, five of the seven randomly selected collector pairs connected by GCL (Figure 4, panels a-d) showed values of $\eta$ that were lower for downstream relative to upstream collectors by a factor of 0.63 ± 0.99. Whereas this discrepancy with experiments is concerning, potential differences in the simulated versus experimental flow fields warrant additional simulation efforts that will be forthcoming.

Mechanistic colloid trajectory simulations in a Happel sphere-in-cell collector under representative unfavorable conditions (Table SI-2) using Parti-Suite freeware Johnson and Pazmino (2023)[4] (https://github.com/drwpjohnson/Parti-Suite) demonstrated approximately factor-of-two decreases in angles of exit versus entry to the fluid envelope and the near-surface fluid domain (defined as 200 nm separation distance) for colloids that intercepted the near surface fluid domain (Figures SI-7, SI-8, Table SI-3). Colloids that exited the fluid envelope without intercepting the near surface fluid domain showed no focusing of exit angles toward the RFSP, corroborating secondary minimum association as the driver of interception focusing (Figures SI-7, SI-8, Table SI-3).

The possibility was also examined in Happel cell simulations that interception focusing increased $\alpha$ via focusing interception more proximal to FFSP where low shear allows increased colloid residence time to bring colloids into contact with heterodomains. This potential impact of interception focusing on $\alpha$ was simulated under the conditions provided above (Table SI-2) using 90 nm radius heterodomains (sufficient to eliminate the repulsive barrier upon complete overlap by colloid ZOI) and varied radii of injection (RLIMs of 1, 4, and 20 $\mu$m) to force interception more proximal to FFSP as RLIM decreased (Figure SI-9). As expected from physicochemical interactions, attachment locations were increasingly distal to the FFSP as RLIM increased and as SCOV decreased; the former involving higher interception angles, the latter involving long near-surface translation distances (Figures SI-9, SI-10).

The Happel-simulated impact on $\alpha$ of interception focusing (decreased RLIM) was inconsistent. For SCOV values of 5E-3 and 5E-5 no significant change in $\alpha$ occurred as RLIM decreased (Figure SI-10). In contrast, for the 5E-4 SCOV condition, an approximately factor-of-two increase in $\alpha$ occurred in response to a factor-of-twenty decrease of RLIM 20 to 1 $\mu$m (Figure SI-9). An additional simulation with RLIM = 100 $\mu$m and SCOV = 5E-5 yielded the same $\alpha$ as RLIM =20 $\mu$m (Figures SI-9 and SI-10, Table SI-4), indicating that $\alpha$ was not impacted by interception focusing at SCOV 5E-5.

*Impact of $\eta$ increase observed in experiments*

The impact on $\alpha$ of interception focusing (decreased RLIM) was non-generalizable. As such, the most apparent impact of interception focusing was the factor of two increase of $\eta$ for downstream relative to upstream collectors across GCL, where were found to be prevalent in the 2D flow cell. Al-Zghoul et al. (2025)[13] developed an expression (Equation 3 below) to predict RPs based on changes in $\alpha$ or $\eta$ with interception order in the context of the interception history framework.

$$N^{att}(x) = N_0 \exp\left(-\frac{k_f}{v}x\right)\sum_{n=1}^{N_i}\left(\frac{\left(\frac{k_f}{v}x\right)^{n-1}}{(n-1)!}\alpha_n\prod_{j=1}^{n}\left(1-\alpha_{j-1}\right)\right) \tag{3}$$

where $n$ is the interception order, and $N_i$ is the maximum number of interception orders considered. Implementing Equation 3 in an Excel spreadsheet (available via link provided in Data Availabiity Statement) with $N_0$ = 2.6E4, $d_c$ = 100 μm, θ = 0.37, $v$ = 4 m/day, $\alpha$ = 0.01 for all interception orders, $\eta$ = 0.08 for the first interception order (single interception attachers), and $\eta$ = 0.16 for interception orders > 1 (multiple interception attachers) yields the result shown in Figure 6 where it is observed that a non-monotonic RP is predicted (as observed in Johnson et al., 2025)[7] for the herein-observed factor-of-two increase in $\eta$ for multiple- relative to single-interception attachers.

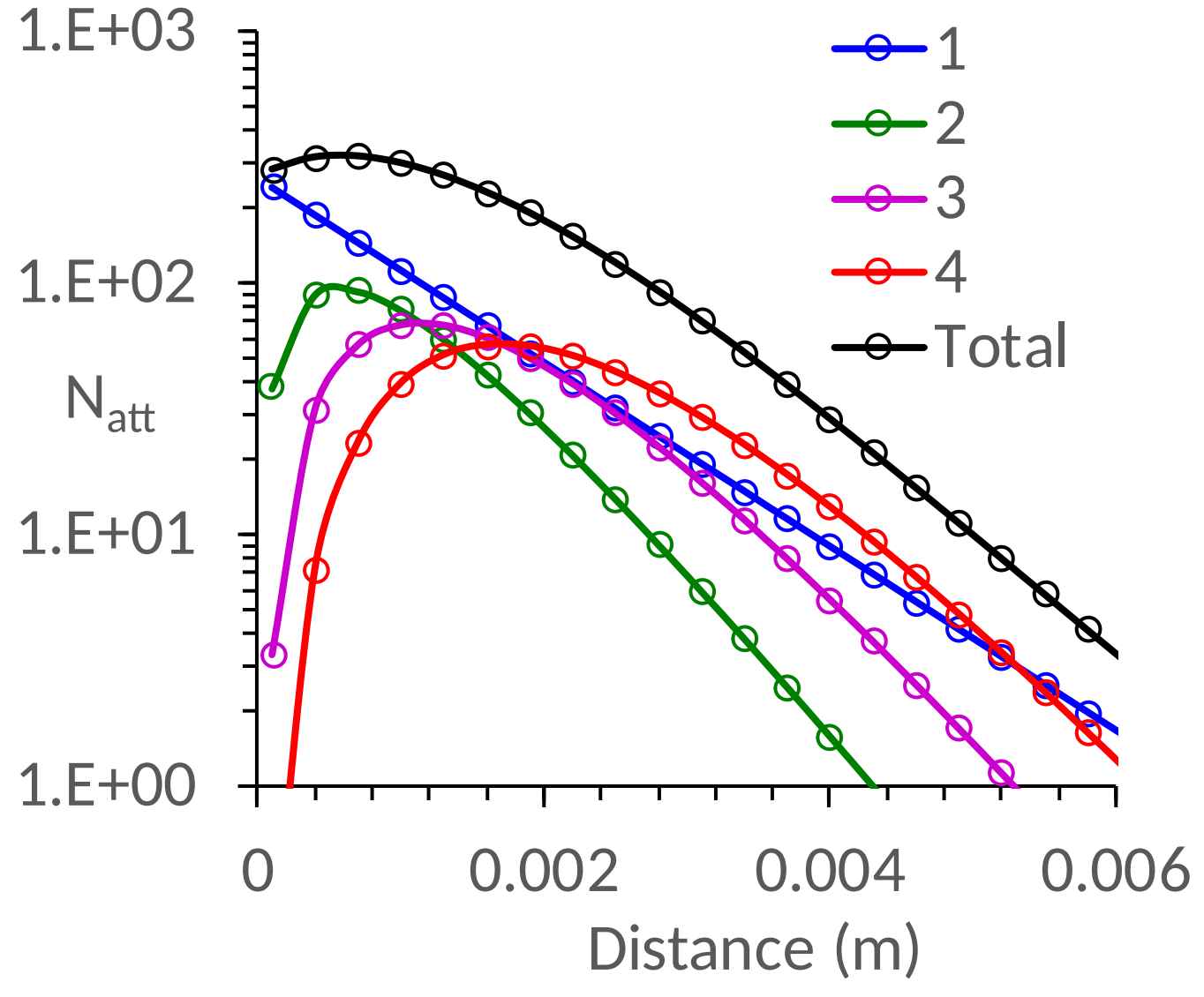

Figure 6. Contributions to Total RP of single- through fourth-order interception RPs according to Equation 3 (Al-Zghoul et al., 2025).[7]

*Direction of* $\alpha$ or $\eta$ *change (how multiexponential RP may be generated)*

We've herein provided evidence that intercepting colloids that fail to attach are focused into a network of ICL that increases their subsequent $\eta$. Whereas increased $\eta$ for subsequent relative to first interceptions explains the onset of nonmonotonic RPs, the onset of multiexponential RPs is not explained by this sole aspect of interception focusing. We speculate below on possible physicochemical mechanisms leading to multiexponential RP while noting that both nonmonotonic and multiexponential RP emerge in a mathematical model accounting for contrasting likelihoods of interception for multiple- versus single-intercepting colloids (Al-Zghoul et al., 2026).[17] One possible mechanism is that $\eta$ may increase or decrease following interception depending on the prevalence of GCL among IML. The number of GCL relative to IML does not seem to change with flowrate since the IML network remains equivalent across factor-of-400 increased flow rate (Figure 1). However, the possibility that GCL prevalence changes with porous media structure, e.g., grain size uniformity or angularity, warrants examination. Likewise, the possibility that decreased $\alpha$ may overwrite increased $\eta$ warrants further examination since blocking of attachment sites (heterodomains) by attached colloids may potentially result from increased focus of colloid interception proximal to the FFSP.

*Some Implications of Interception Focusing and Interception History*

Despite its highly transient nature for solutes, incomplete pore scale mixing decreases solute reaction yields during reaction in granular porous media (e.g., de Anna et al., 2013; Gramling et al., 2002).[8,18] The contrastingly persistent nature of incomplete pore scale mixing for colloids was herein shown to convey colloids expelled proximal to RFSP (multiple interceptors) to increased likelihood of subsequent interception as suggested by decreased transport distance between interceptions for multiple interceptors (Figure 2) as driven by expulsion proximal to RFSP for multiple interceptors (Figure 3), increased $\eta$ for downstream relative to upstream collectors linked proximally by GCL (Figure 4), and a significant prevalence of GCL among ICL (Figure 5). While each of these elements warrant further definition, in aggregate they provide strong support that interception focusing is an inevitable outcome of relatively low diffusivity, relatively long range secondary minimum attraction, and persistent incomplete pore scale mixing for colloids. Interception focusing was demonstrated in both experiments and simulations at the grain scale.

Extended tailing of low colloid concentrations following initial elution that is observed in colloid breakthrough-elution concentration histories (BTECs) under unfavorable conditions (e.g., Johnson et al., 1995; Johnson and Hilpert, 2013)[19,20] emerges in a mathematical model that accounts for contrasting likelihoods of interception for multiple- versus single-intercepting colloids (Al-Zghoul et al., 2026).[17] Researchers often attempt to describe extended tailing using a first-order detachment rate coefficient in models such as Hydrus (Šimůnek et al., 2016)[21] to generate return of colloids from grain surfaces to bulk fluid. However, this approach is inconsistent with the fact that colloid attachment is largely irreversible in the absence of perturbations in flow and solution chemistry (Pazmino et al., 2014b; Rasmuson et al., 2019; VanNess et al., 2019),[1-3] and more relevant to this paper, is inconsistent with the non-Markovian nature of extended tailing (Ginn, 2000)[22] reflecting "memory" wherein retention of a colloid begets its further retention (Johnson et al., 1995).[19] This inconsistency holds also when Hydrus forces the attachment rate coefficient to decrease with depth to produce multiexponential RPs (so-called straining) or with attachment (blocking) to produce a transient rise in the BTEC plateau associated with step injection. Interception focusing provides a physicochemical basis for the non-Markovian nature of extended tailing, and suggests a relationship between RP shape and extended tailing that is linked to changes in $\eta$ and possibly $\alpha$ with interception order.

Both nonmonotonic and multiexponential RP demonstrate pre-asymptotic initial portions that extend across a finite distance in the RP (Figure SI-2). This pre-asymptotic distance was observed to increase with grain size for experimentally observed multiexponential RP (Johnson et al., 2020).[23] We posit that this pre-asymptotic distance corresponds to the distance across which IML (and GCL) prevalence increases to a steady state; i.e., IML prevalence increases from the inlet via their emanation from grain RFSPs, and IML prevalence reaches a steady state via IML termination via their merging and via interception of grain FFSP (GCL). The possibility that pre-asymptotic portions of RPs reflect IML and GCL development warrants examination in future work.

We stress that the "anomalies" of nonexponential RP and extended tailing are observed under unfavorable conditions (and are absent under favorable conditions) under the simplest possible physical conditions, i.e., uniform colloids and uniform grains under laminar flow (Johnson and Pazmino, 2023),[4] as demonstrated in Li et al. (2004), Li and Johnson (2005), and Tong and Johnson (2006),[24-26] which RP and BTEC data are available via a link provided in the Data Availability Statement. The inevitable and under-appreciated impacts of low diffusivity, long range secondary minimum attraction, and temporally persistent incomplete pore scale mixing for colloids constitute a fundamental distinction from solutes that explains ubiquitous "anomalies" associated with colloid transport. To put it colloquially, these so-called "anomalies" are a "feature" rather than a "bug" of colloid transport under unfavorable conditions.

**Supporting Information**: 20 pages, 5 tables, and 10 figures and text providing derivation of Equation 2.

**Data Availability**

The data and code used for the simulations and experiments described are publicly available online at:

https://doi.org/10.5281/zenodo.20598400 and http://dx.doi.org/10.5281/zenodo.15319454 and from the corresponding author on request.

**Author Contributions**

W.P.J, and D.B contributed to the study concept. N.W. and L.A.U. carried out experiments. N.W., W.P.J. and B.A. carried out simulations. W.P.J., N.G.W., and L.A.U. carried out data analysis. W.P.J and N.W.G. drafted the manuscript. All co-authors edited the manuscript and approved the final, submitted manuscript.

**Acknowledgements**

The authors wish to thank three anonymous reviewers for their helpful comments. This material is based upon work supported by grants from the US National Science Foundation WALCZ program EAR-2538202 and EAR-2538203, and from the Hydrologic Sciences program EAR-1951676, and EAR-1951677. Any opinions, findings, and conclusions or recommendations expressed in this material are those of the authors and do not necessarily reflect the views of the National Science Foundation. Support for parallel computation utilized in Happel transport simulations was provided by the University of Utah Center for High Performance Computing.

Supporting Information for:

# Interception Focusing of Colloid Transport in Granular Porous Media

William P. Johnson[1,*] Noah G. Willis[1], Luis A. Ullauri[1], Bashar M. Al-Zghoul[2], Diogo Bolster[3]

[1]University of Utah, Geology and Geophysics, Salt Lake City, Utah 84112, USA

[2] University of California Davis, Civil and Environ. Eng., Davis, California 95616, USA

[3]University of Notre Dame, Civil and Environmental Engineering and Earth Sciences, Notre Dame, Indiana 46556, United States

**Contents of this file: 22 pages, 5 Tables, 10 Figures**

Figure SI-6. Vector plot of Happel simulated colloid trajectory locations and corresponding mean angles for entry to, and exit from, the fluid envelope and near surface fluid domain.

Figure SI-7. Box plot of Happel simulated colloid trajectory locations and corresponding mean angles for entry to, and exit from, the fluid envelope and near surface fluid domain.

Figure SI-8. Location plot in z-r plane of Happel sphere-in-cell simulated outcomes under unfavorable conditions with varied SCOV and RLIM under conditions in Table SI-1.

Figure SI-9 Location plot in x-y plane of Happel sphere-in-cell simulated outcomes under unfavorable conditions.

Figure SI-10 Happel sphere-in-cell simulated values of $\alpha$ as a function of limiting radius of injection (RLIM) and surface coverage by heterodomains (SCOV).

## Relationship between $k_f$ and $\eta$ in 2D flow cell

The relationship between $k_f$ and $\eta$ is as follows (e.g., Johnson and Pazmino, 2023):

$$k_f = -\frac{N_c}{L} v \ln(1-\eta) \qquad \text{(SI-1)}$$

where $v$ is the average fluid velocity in the fluid domain.

In a control volume of dimension $\Delta x \Delta y \Delta z$, with a number of cylindrical collectors per control volume ($N^*_c$):

$$N_c^* \Delta x \Delta y \Delta z = \frac{\Delta x \Delta y \Delta z (1-\varepsilon_{bulk})}{\Delta z \frac{\pi}{4} d_{ns}^2} \qquad \text{(SI-2)}$$

where $\varepsilon_{bulk}$ is the porosity defined by considering $d_{ns}$ as the diameter corresponding to the near-surface "collector" (radius of solid collector plus near-surface fluid thickness).

Or:

$$N_c^* = \frac{(1-\varepsilon_{bulk})}{\Delta z \frac{\pi}{4} d_{ns}^2} \qquad \text{(SI-3)}$$

Multiplying by the area of the unit collector ($\Delta z b$) yields the number of spherical collectors per length ($N_c/L$):

$$\frac{N_c}{L} = \frac{(1-\varepsilon_{bulk}) b}{\frac{\pi}{4} d_{ns}^2} \qquad \text{(SI-4)}$$

For the cylindrical unit collector, the fluid shell diameter ($b$) relative to the near-surface "collector" diameter ($d_{ns}$) preserves the bulk porosity, yielding the relationship:

$$\frac{d_{ns}}{b} = (1-\varepsilon_{bulk})^{\frac{1}{2}} \qquad \text{(SI-5)}$$

Substituting the above into the expression for $N_c/L$ yields:

$$\frac{N_c}{L} = \frac{4\left(1 - \varepsilon_{bulk}\right)^{\frac{1}{2}}}{\pi\, d_{ns}} \qquad \text{(SI-6)}$$

Substituting the above into Equation SI-1 yields kf for the 2D porous media:

$$k_f = -\frac{4\left(1 - \varepsilon_{bulk}\right)^{\frac{1}{2}}}{\pi\, d_{ns}} v \ln\left(1 - \eta\right) \qquad \text{(SI-7)}$$

**Table SI-1**. Values of $\eta$ and $\alpha$ for downstream relative to upstream collectors connected by GCL. Identification of pairs and triplets correspond to the panels of Figure 4. Averages and standard deviations correspond to all pairs (including within triplet) or to proximal pairs (excepting the first pair in triplet 2 of panel (e).

| Panel | ID | $\eta$ | $\alpha$ | Ratios downstream/upstream | |
|---|---|---|---|---|---|
| a | triplet 1 | 0.048 | 0.067 | $\eta$ | |
| a | triplet 1 | 0.09 | 0 | 1.9 | |
| a | triplet 1 | 0.12 | 0.024 | 1.3 | |
| b | pair 1 | 0.081 | 0.03 | | |
| b | pair 1 | 0.16 | 0.015 | 2.0 | |
| c | pair 2 | 0.062 | 0.038 | | |
| c | pair 2 | 0.15 | 0.018 | 2.4 | |
| d | pair 3 | 0.14 | 0.091 | | |
| d | pair 3 | 0.22 | 0.048 | 1.6 | |
| e | triplet 2 | 0.069 | 0.043 | | |
| e | triplet 2 | 0.046 | 0 | 0.7 | |
| e | triplet 2 | 0.22 | 0.048 | 4.8 | |
| | | | | | |
| | | All | average | 2.1 | |
| | | All | Std dev | 1.2 | |
| | | Proximal | average | 2.3 | |
| | | Proximal | Std dev | 1.1 | |

**Table SI-2**. Input parameters for trajectory simulations in Happel sphere-in-cell collector

********Simulation parameters (integer values)

| # of colloids | Perfect sink(0) or Contact(1) or Perturbation(-1) mode | | Single(0) or Parallel(1) simulat. |
|---|---|---|---|
| NPART | HNS | ATTMODE | CLUSTER |
| 5 | 2.0E-7 | 1 | 1 |

********Flow and simulation geometry parameters (Continuum flow field applies for VJET<1.273 m/s)

| Fluid velocity | Injection radius | Happel porosity | Happel Collector Radius | Total simul. |
|---|---|---|---|---|
| VSUP(m/s) | RLIM(m) | POROSITY | AG(m) | TTIME(s) |
| 1.620E-4 | 20.0E-06 | 0.35 | 2.550E-04 | 5000.0 |

********Colloid and fluid properties

| Colloid radius | Colloid density | Fluid density | Fluid viscosity | Relative permittivity | Temperature |
|---|---|---|---|---|---|
| AP(m) | RHOP(kg/m3) | RHOW(kg/m3) | VISC(kg/m/s) | ER(-) | T(K) |
| 5.0E-07 | 1.050E+03 | 9.980E+02 | 9.980E-04 | 8.000E+01 | 293.2 |

********Water chemistry and surface charge parameters

| Ionic strength | Valence of electrolyte | Colloid zeta potential | Collector zeta potential |
|---|---|---|---|
| IS(mol/m3) | ZI(-) | ZETAPST(V) | ZETACST(V) |
| 6.00E+00 | 1.0 | -6.50E-02 | -6.50E-02 |

********Heterogeneity parameters collector (SCOV to 0 for no heterodomains, HETMODE 1 for uniform, 9 for bimodal 1:8 large: small)

| Heterodomain ζ | HETMODE | Large het. rad. | Small het. rad. | Fractional surface coverage |
|---|---|---|---|---|
| ZETAHET(V) | HETMODE | RHET0(m) | RHET1(m) | SCOV(-) |
| 6.500E-02 | 1 | 9.000E-08 | 0.000E+00 | 5.0E-5 |

********Heterogeneity parameters colloid (SCOVP to 0 for no heterodomains, HETMODEP to 1 for uniform, 9 for bimodal 1:8 large:small)

| Heterodomain ζ | HETMODEP | Large het. rad. | Small het. rad. | Fractional surface coverage |
|---|---|---|---|---|
| ZETAHETP(V) | HETMODEP | RHETP0(m) | RHETP1(m) | SCOVP(-) |
| 6.500E-02 | 1 | 0.000E+00 | 0.000E+00 | 0.000E+00 |

********Van der Waals force parameters - Make sure to set VDWMODE (If VDWMODE = 2,3,4; A132(J) entry is inconsequential) Uncoated (1), Layered colloid - Layered collector (2), Colloid - Layered collector (3), Layered colloid - collector (4)

| Combined Hamaker constant | vdW decay length | VDWMODE |
|---|---|---|
| A132(J) | LAMBDAVDW(m) | VDWMODE |
| 7.18E-21 | 100.0E-9 | 1 |

********Van der Waals force parameters - Coated systems – H = Hamaker Constant

| Colloid H | Colloid coating H | Collector H | Collector coating H | Fluid H |
|---|---|---|---|---|
| A11(J) | AC1C1(J) | A22(J) | AC2C2(J) | A33(J) |
| 0.0 | 0.0 | 0.0 | 0.0 | 0.0 |

********Van der Waals force parameters - Coated systems

| Colloid coating thickness | Collector coating thickness |
|---|---|
| T1(m) | T2(m) |
| 0.0 | 0.0 |

********Lewis acid-base and steric hydration force parameters

| AB energy at min. sep. | AB decay length | Ste. energy at min. sep. | Ste. decay length |
|---|---|---|---|
| GAMMA0AB(J/m2) | LAMBDAAB(m) | GAMMA0STE(J/m2) | LAMBDASTE(m) |
| -2.7E-2 | 6.0E-10 | 1.700E-2 | 4.100E-10 |

********Roughness parameters

(Set Roughness mode to 0 for no roughness, 1 for rough colloid, 2 for rough collector, and 3 for rough colloid & collector)

(For Max Asperity Height coarse scale for lever arm, assume = B/2 to be consistent with Matlab version)

| Slip Length | Roughness mode | Asperity Height (fine scale) | Max Asperity Height |
|---|---|---|---|
| B(m) | RMODE (-) | ASP (m) | ASP2 (m) |
| 0.000E+00 | 0 | 0.000E+00 | 0.0E-9 |

********Deformation parameters (W132 is negative for attractive)

| Combined elastic modulus | Work of Adhesion | Contact Radius Factor (1.0 for detachment, <1.0 for attachment) |
|---|---|---|
| KINT(N/m2) | W132(J/m2) | BETA(-) |
| 4.36E9 | -2.9E-2 | 0.5 |

********Diffusion and Gravity factors

(DIFFSCALE calibrated to 1.35 or 0.0 for none, GRAVFACT is factor of gravity acceleration (1 for Earth, 0.0 for none)

| Multiplier of diffusion force | Multiplier of gravity force | Direction of Gravity 1=with flow, 2=against flow, 3=orthog. to flow in -X dir., 4=orthog. to flow in +X dir) |
|---|---|---|
| DIFFSCALE(-) | GRAVFACT(-) | CBGRAV(-) |
| 1.35 | 1.0 | 1 |

********Simulation time step multipliers and distance displacement factors

Momentum relaxation time (MRT) multipliers and Factors to compare distance expected from diffusion

| Mult. Bulk | Mult Near-surf | Mult. Contact | Fact. Near-surf. | Fact. Contact |
|---|---|---|---|---|
| MULTB(-) | MULTNS(-) | MULTC(-) | DFACTNS(-) | DFACTC(-) |
| 1.000E+02 | 2.000E+00 | 0.01 | 1.0E-4 | 1.0E-5 |

********Output and Detachment parameters (integers)

| Output interval | Maximum output array size |
|---|---|
| NOUT | PRINTMAX |
| 1000 | 10000 |

**Table SI-3**. Happel sphere-in-cell collector simulated colloid trajectory mean, median, and standard deviation (population) angles relative to z-axis (RFSP pole) for entry and exit to/from the fluid envelope (Ent-Rb-z and Ex-Rb-z, respectively) and to/from the near surface fluid domain (Ent-NS-z and Ex-NS-z, respectively) for the attached (Att), remained without attachment (Rem), exited with and without visit to the near surface fluid domain (ExitNS and Exit, respectively). Simulations were run with parameters in Table SI-2.

| Angles to z | | Att | Rem | ExitNS | Exit |
|---|---|---|---|---|---|
| Ent-Rb-z | Ave | 2.78E+00 | 3.36E+00 | 3.46E+00 | 1.35E+01 |
| Ent-Rb-z | Med | 2.68E+00 | 3.07E+00 | 3.32E+00 | 1.42E+01 |
| Ent-Rb-z | StdDevP | 1.39E+00 | 1.82E+00 | 1.56E+00 | 4.37E+00 |
| Ex-Rb-z | Ave | | 2.51E+00 | 1.49E+00 | 1.37E+01 |
| Ex-Rb-z | Med | | 2.53E+00 | 8.88E-01 | 1.42E+01 |
| Ex-Rb-z | StdDevP | | 1.46E-01 | 1.45E+00 | 4.59E+00 |
| Ent-NS-z | Ave | 5.05E+01 | 5.87E+01 | 6.12E+01 | |
| Ent-NS-z | Med | 5.18E+01 | 6.51E+01 | 6.49E+01 | |
| Ent-NS-z | StdDevP | 2.43E+01 | 2.28E+01 | 1.95E+01 | |
| Ex-NS-z | Ave | | | 2.98E+01 | |
| Ex-NS-z | Med | | | 1.95E+01 | |
| Ex-NS-z | StdDevP | | | 2.53E+01 | |

**Table SI-4**. Tabulated attached, remained without attachment, and exit results from Traj-Hap simulations. 5000 colloids with RLIM = 100 μm and SCOV = 5E-5 with the parameters provided in Table SI-2.

| | Att | Rem | ExitNS | Exit | Total | # approach | 4.33E+04 |
|---|---|---|---|---|---|---|---|
| # | 8 | 57 | 115 | 4820 | 5000 | 0.0042 | $\eta$ |
| Interceptions | 8 | 57 | 115 | 0 | 180 | 0.036 | $\eta^*$ |
| Attachments | 8 | 0 | 0 | 0 | 8 | 0.044 | $\alpha$ |

**Table SI-5**. Tabulated F-I angles for select upstream versus downstream collectors across GCL.

| Panel | Collector 1 | Collector 2 | Collector 3 |
|---|---|---|---|
| A | 89.34 | | 30.44 |
| B | 11.65 | 10.8 | |
| C | 15.97 | 89.9 | |
| D | 25.18 | 1.8 | |
| E | 6.18 | 169.15 | 4.01 |

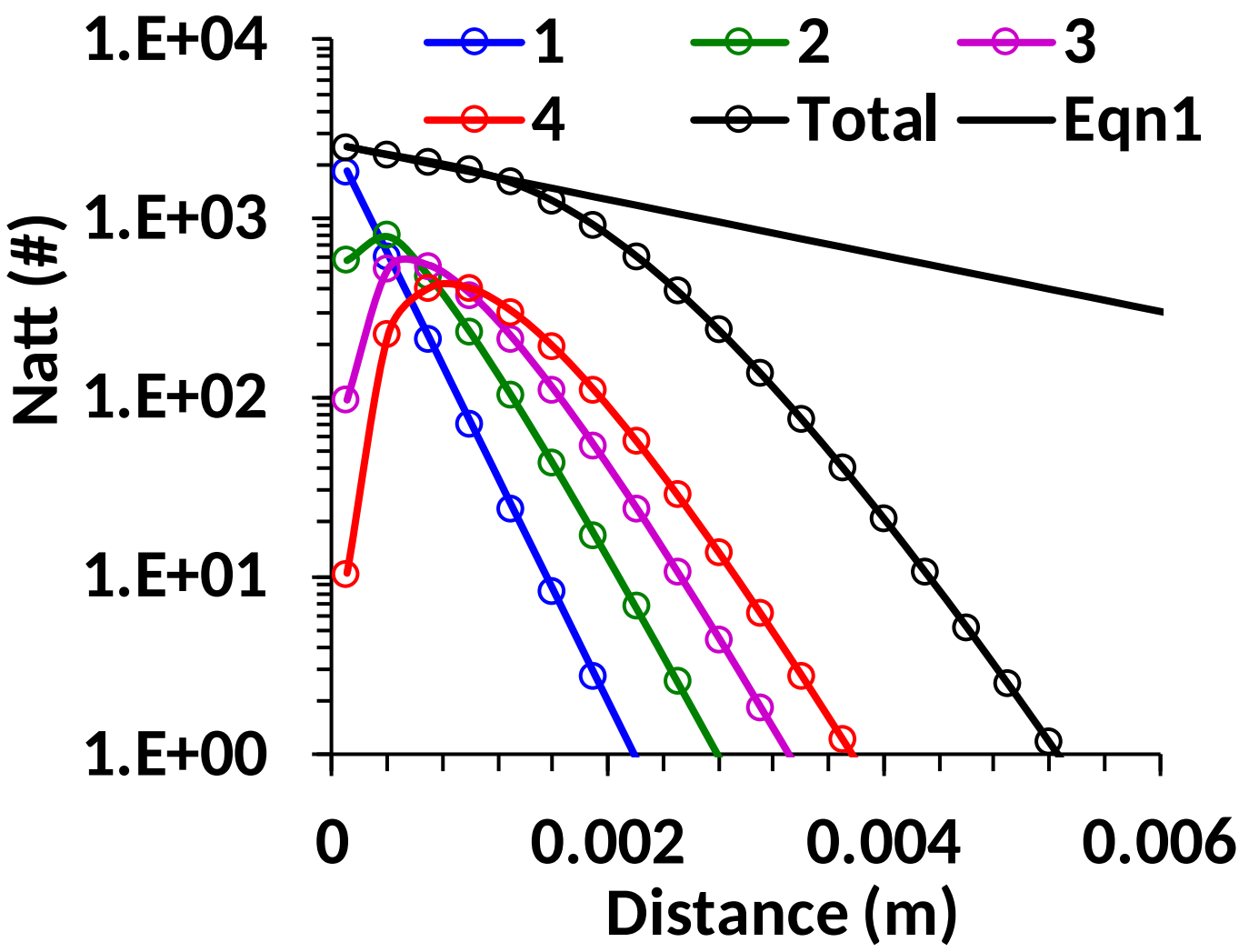

**Figure SI-1**. RPs computed in 2D flow cell using Equations 1 and 2 (line with no symbols, Eqn1) and computed using Equations 3 and 2 (lines with symbols showing RPs for interception orders 1 through 4 and Total being the total superimposed through 7 interception orders) each assuming no change in $\eta$ or $\alpha$ with increased interception order and using the following values: $N_0$ = 2.6E+04, $d_c$ = 1.00E-04 m, $\theta$ = 0.37, $v$ = 4 m/day, $\eta$ = 0.3, $k_f$ = 1.44E+04/day, $\alpha$ = 0.1.

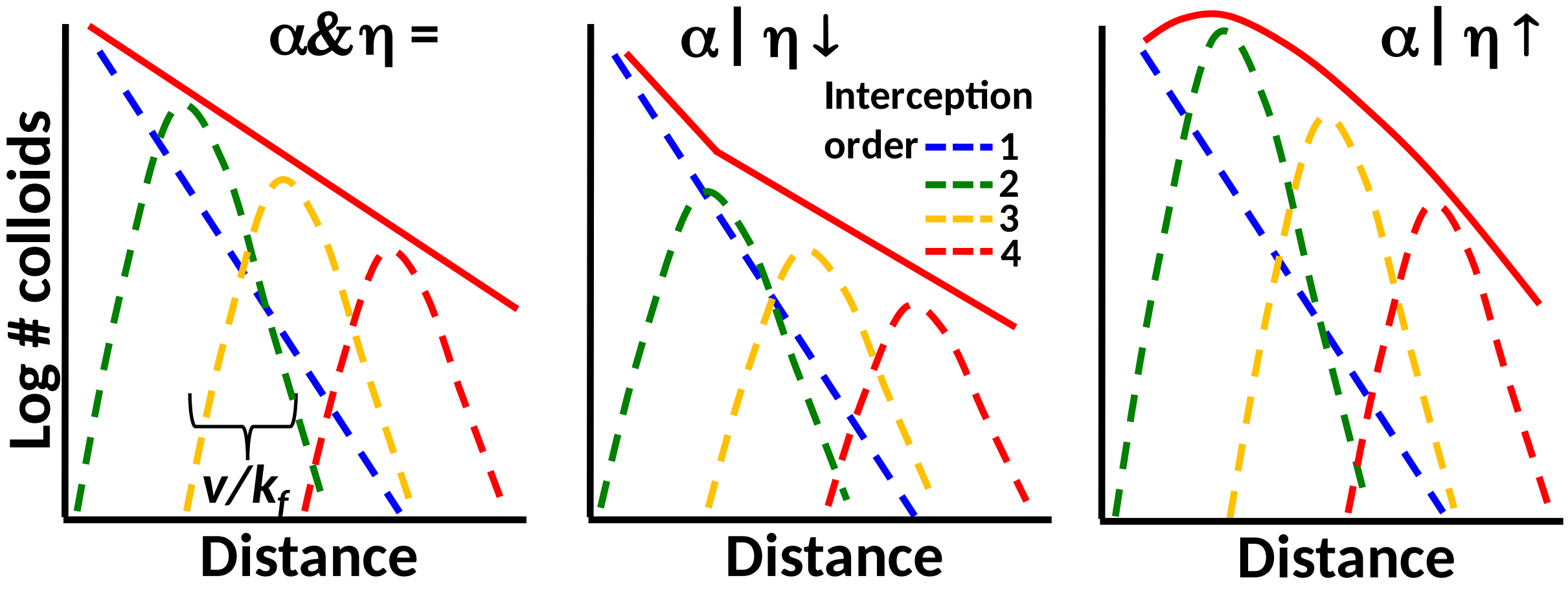

**Figure SI-2**. Schematic of RP shapes based on relationship of $\alpha$ and $\eta$ for multiple versus single-intercepting attachers wherein exponential RP (left panel) reflect equal $\alpha$ and $\eta$ for multiple- versus single-intercepting attachers, multiexponential RP (middle panel) reflect decreased $\alpha$ or $\eta$ for multiple- versus single-intercepting attachers, and nonmonotonic RP (right) reflect increased $\alpha$ or $\eta$ for single- versus multiple-intercepting attachers. The overall RP is denoted by a solid red line, whereas dashed lines indicate single and multiple intercepting attacher RPs, with increasing color warmth corresponding to increased interception order. The average pore water velocity ($v$) relative to the interception rate coefficient ($k_f$) defines a characteristic spacing between peaks in the RPs for multiple interception attachers. After Johnson et al. (2025).

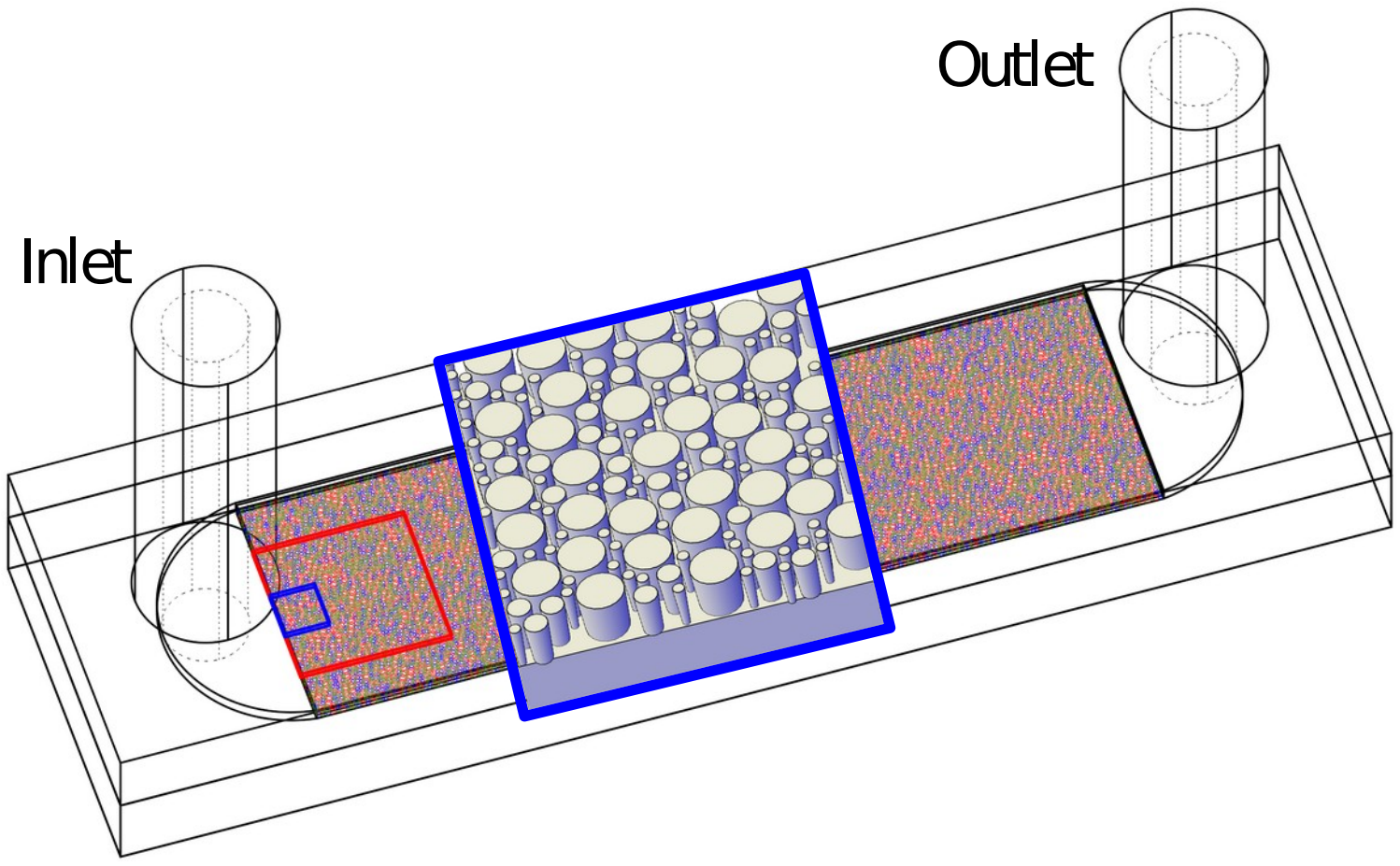

**Figure SI-3**. Schematic of the 2D flow cell showing the chamber with porous media (colorful rectangle), and inlet and outlet for colloidal injection. The field of view under experimental conditions is highlighted in the red square. A magnified view of the porous media is depicted in the blue square.

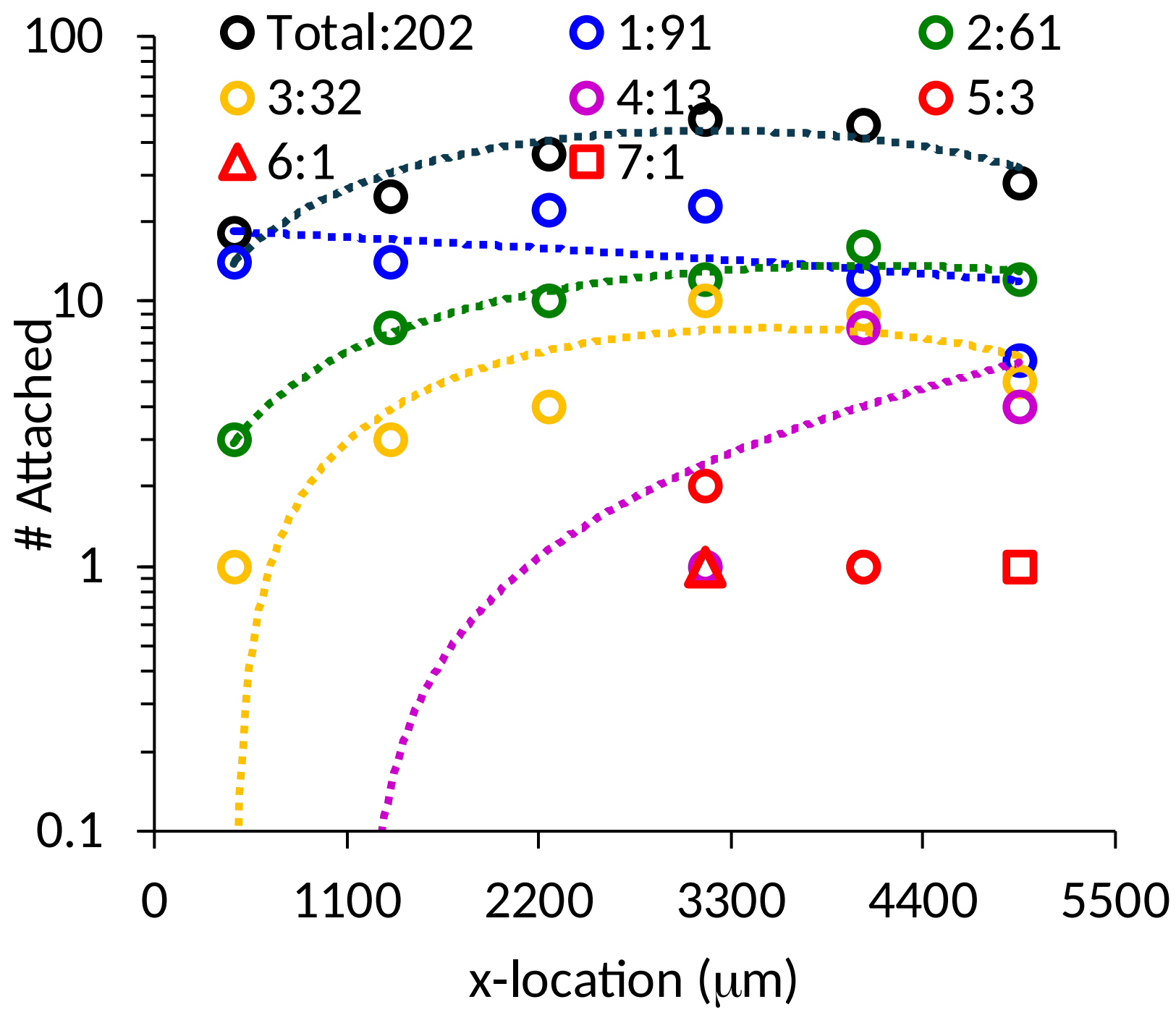

**Figure SI-4**. Number of attached colloids versus transport distance in the field of view under unfavorable conditions in experiments. Symbol and line colors correspond to interception order, with warm colors denoting higher number of interceptions prior to attachment. Legends in show the number of attachments associated with each interception order as interception order:attachments corresponding to that interception order. Exponential and quadratic trend lines (dashed) were fit to single- and multiple-interception data, respectively, for the purpose of associating data according to interception order.

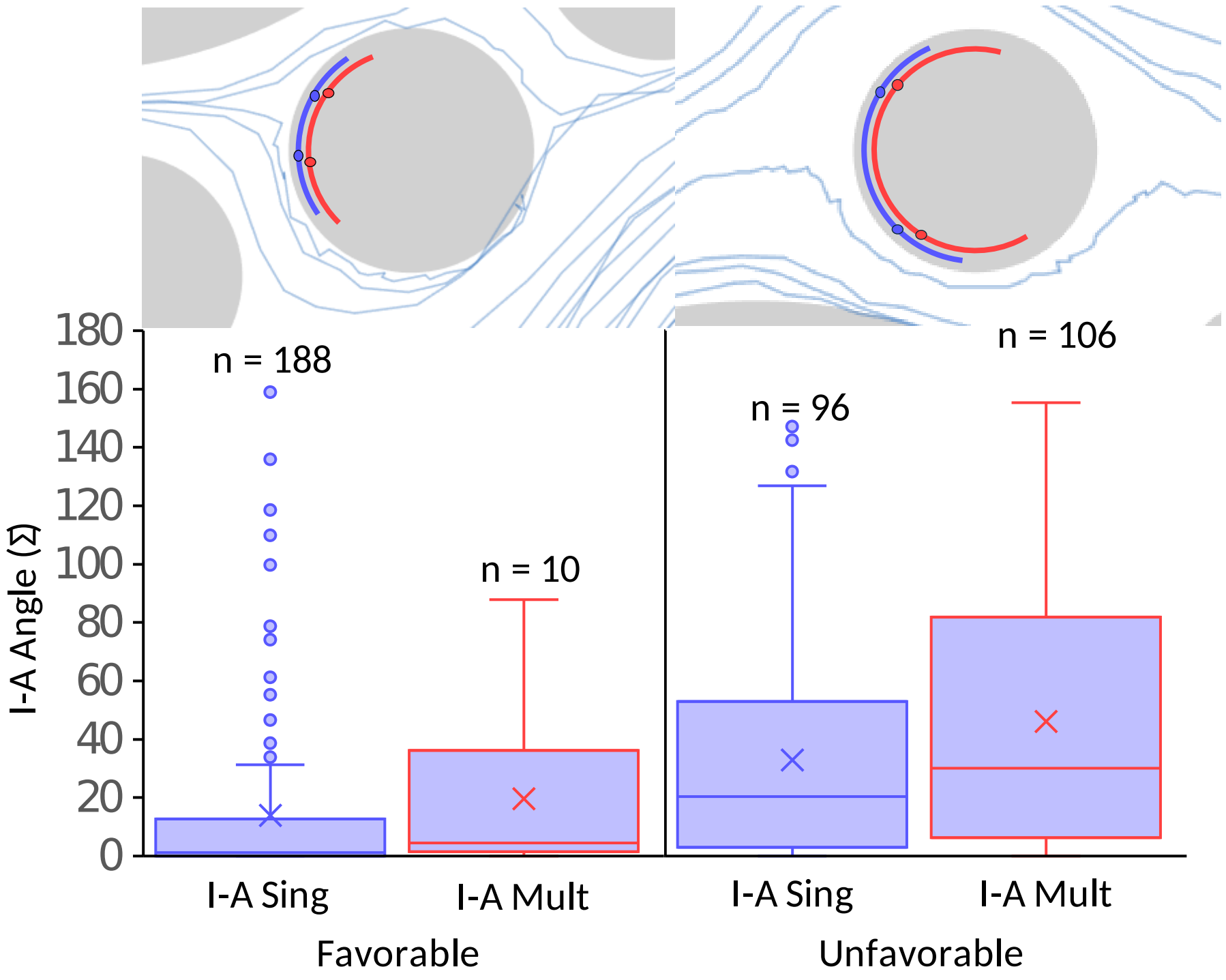

**Figure SI-5**. Interception-to-attachment (I-A) arcs and angles for single and multiple intercepting attachers across favorable (left panels) and unfavorable (right panels) conditions. Top panels: Superimposed example colloid trajectories (blue lines) in a representative portion of the 2D flow cell. Circles and arcs on grain (cylinder) peripheries indicate mean and standard deviation, respectively, of single- (blue) and multiple- (red) intercepting attachers and are duplicated on each side of the FFSP. Bottom panels: Boxplots of I-A angles (degrees) for attached colloids for single- (blue outline) and multiple- (red outline) interceptors. Generated from script for Figure 3.

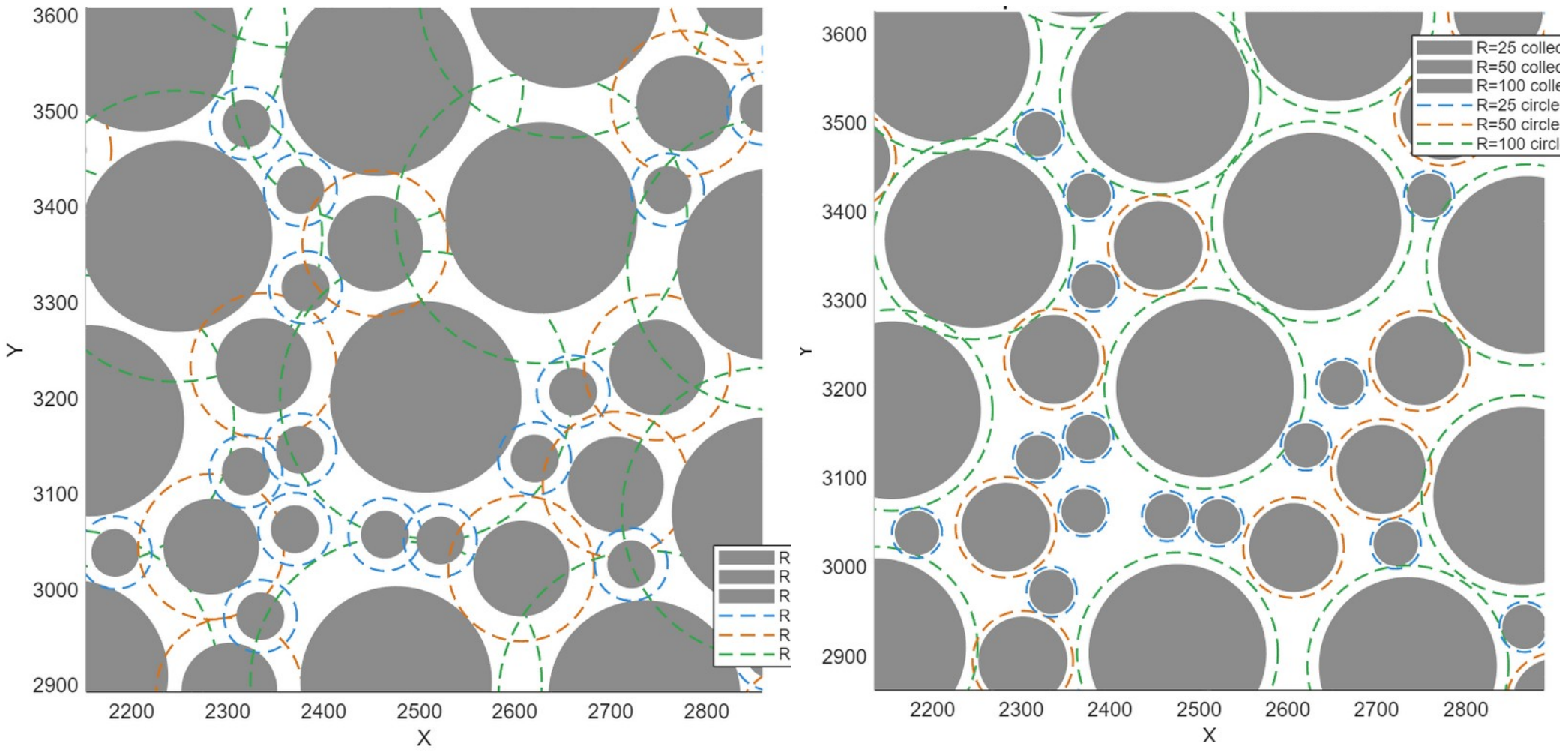

**Figure SI-6**. Fluid shells (radius $R_b$) surrounding grains (cylinders) (radius $R_g$) determined from relationships $R_b/R_g = (1-\theta)^{1/2}$ (left panel) and $R_b/R_g = 0.25*(1-\theta)^{1/2}$ (right panel). Generated using script for Figure 4.

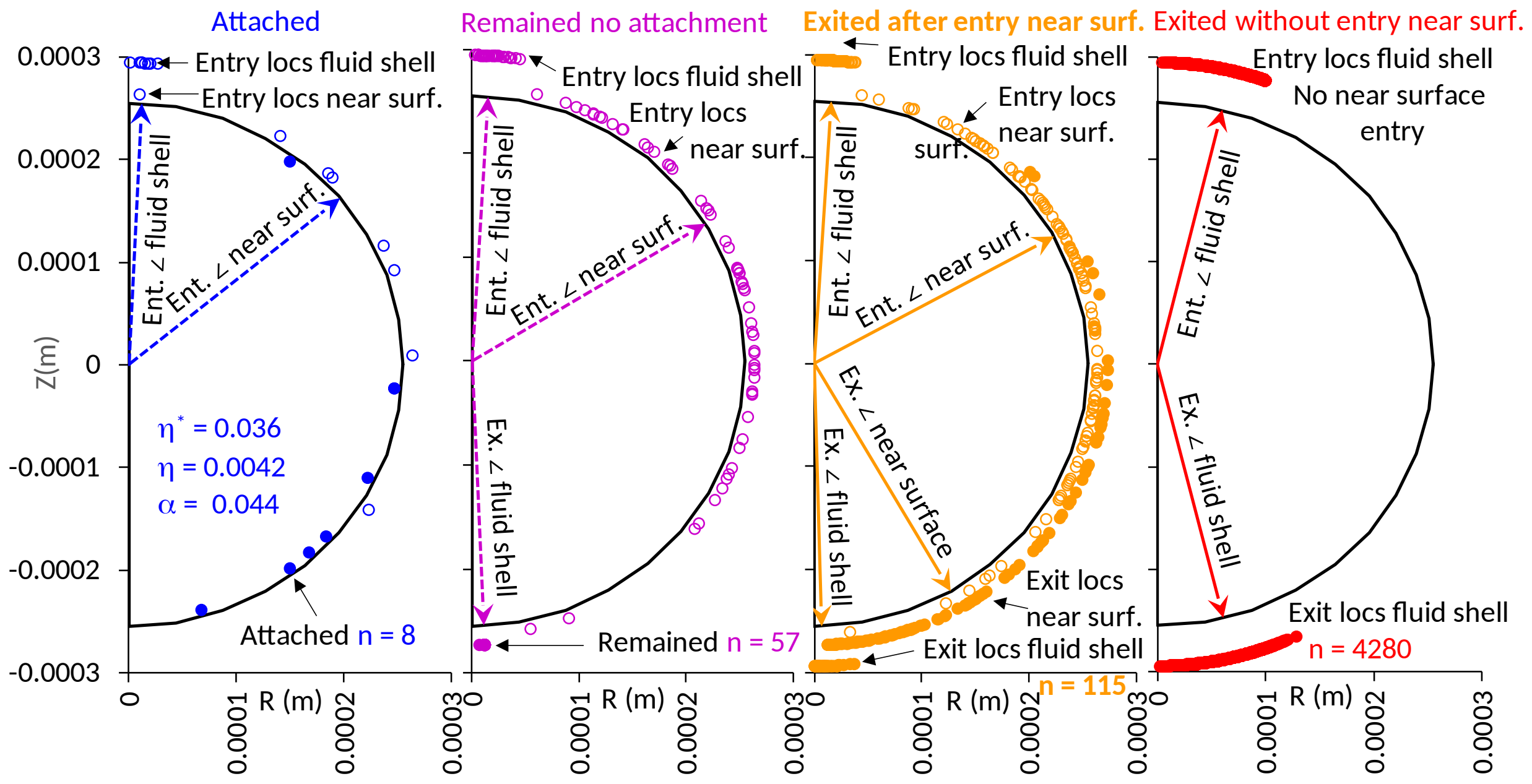

**Figure SI-7**. Simulated colloid trajectory locations and corresponding mean angles for entry to, and exit from, the fluid envelope (shell) and near surface fluid domain in a Happel sphere-in-cell collector with parameters in Table SI-2. Locations and mean angles are collapsed onto a single Z-R plane where flow was from positive to negative Z. Separate panels show locations for colloids that attached (left-most panel), colloids that remained without attachment (middle-left panel), colloids exited after entering near surface fluid domain (middle-right panel), and colloids that exited without entering the near surface fluid domain (right-most panel).

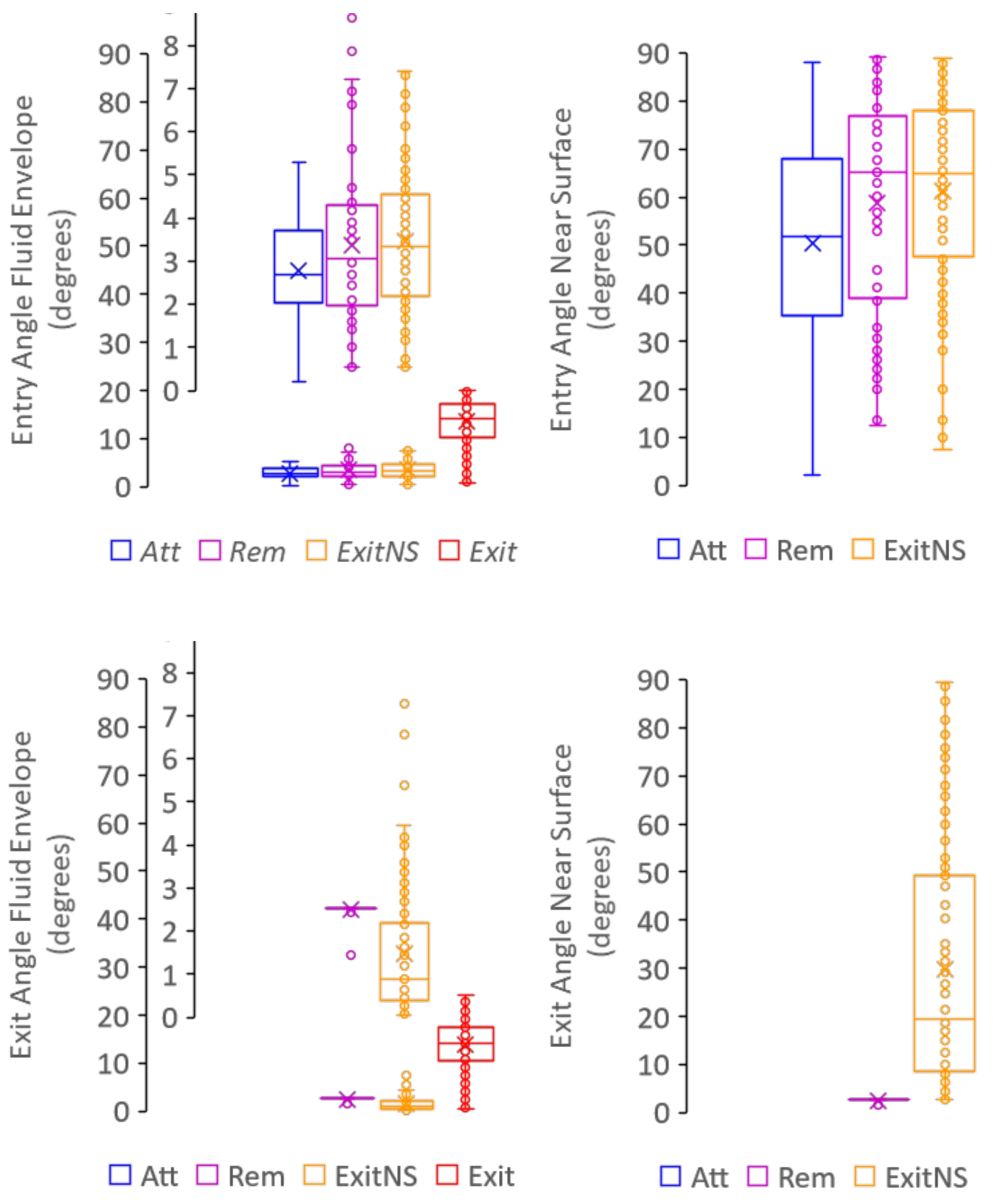

**Figure SI-8**. Happel sphere-in-cell collector simulated colloid trajectory angles for entry (top) and exit (bottom) angles to the fluid envelope (left) and near surface fluid domain (right) for the subpopulations that attached (Att, blue), remained without attachment at RFSP (Rem, magenta), exited subsequent to intercepting the near surface fluid domain (ExitNS, gold), and exited without intercepting near surface fluid domain (Exit, red).

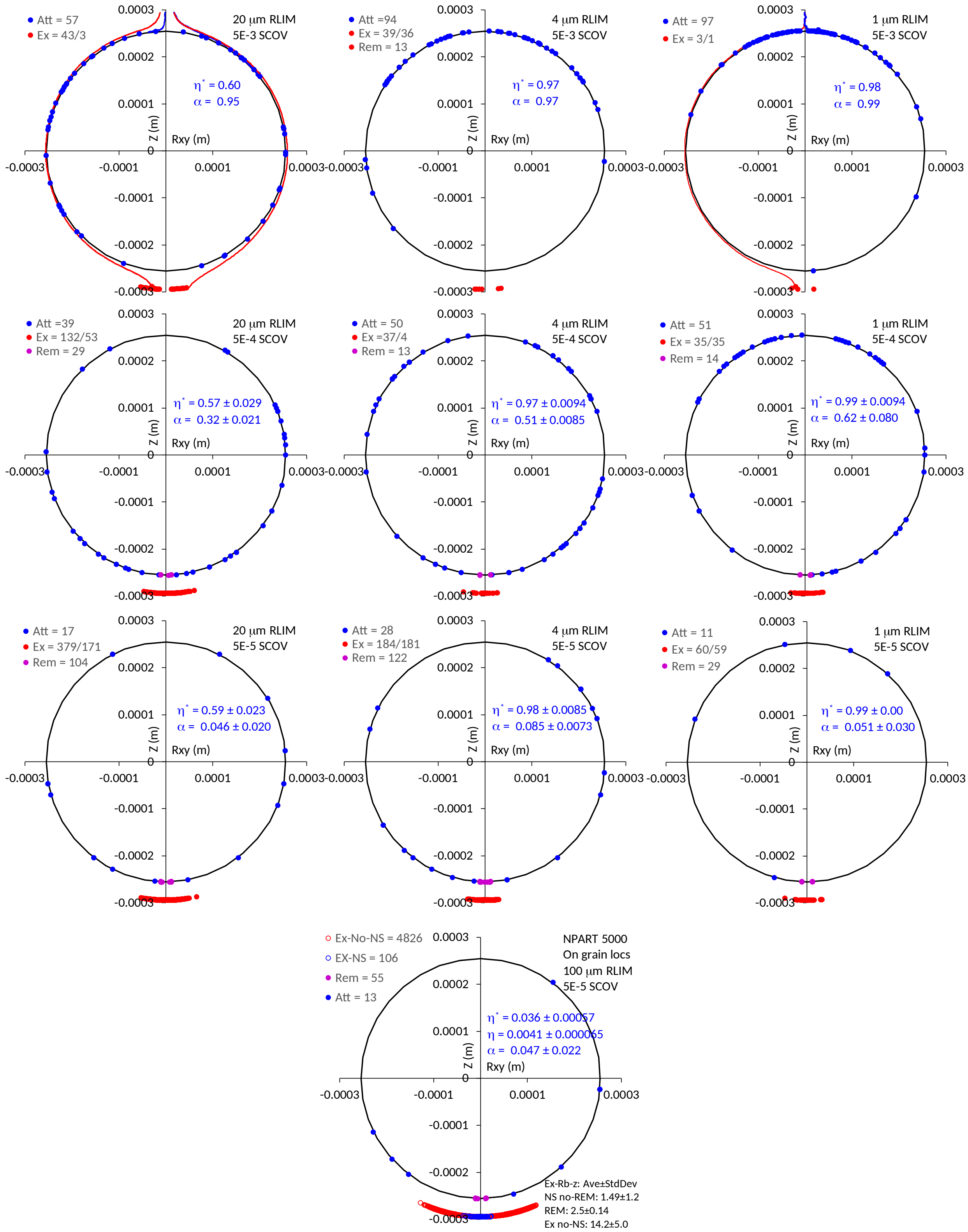

**Figure SI-9**: Simulations for 100 colloids showing locations and numbers of attached (Att = blue dots), remained without attachment (Rem = magenta dots), and exited the fluid envelope (Ex = red dots) with the number of intercepting colloids shown after the / symbol for the latter. Coordinates in x-y are collapsed onto a single axis (Rxy), and flow is downward from positive z. Select trajectories are shown for attached (blue lines) and exited trajectories (red lines). The corresponding SCOV (according to row) and RLIM (according to column). Simulations involved default parameters from Parti-Suite for CML except the colloid diameter was 1-$\mu$m, RHET was 90 nm (Table SI-2). Colloids remaining without attachment (non-contact) were not included as attached in calculation of $\alpha$, since they were assumed to expel to bulk fluid as observed in experiments (Johnson and Pazmino, 2023). For all simulations, less than 1% of attached colloids were within one radius of one other, indicating that the lack of accounting for potential blocking of attachment by attached colloids in simulations had negligible impact. For a simulation conducted with RLIM = 100 $\mu$m, only 8 out of 5000 simulated attached out of the 180 colloids that intercepted the near surface fluid domain (Table SI-3), producing values of $\eta$ = 0.0042 and $\alpha$ = 0.044, where the provisional value of $\eta^*$ (=0.036) for interception of 180 colloids out of 5000 simulated colloids for RLIM 100 $\mu$m was upscaled to 180 interceptions out of 43,300 colloids that would have approached the collector (grain plus fluid envelope = 294 $\mu$m radius) (Table SI-4), as justified by the fact that colloids introduced at radii greater than RLIM had zero chance of intercepting the near surface fluid domain (Figure SI-10).

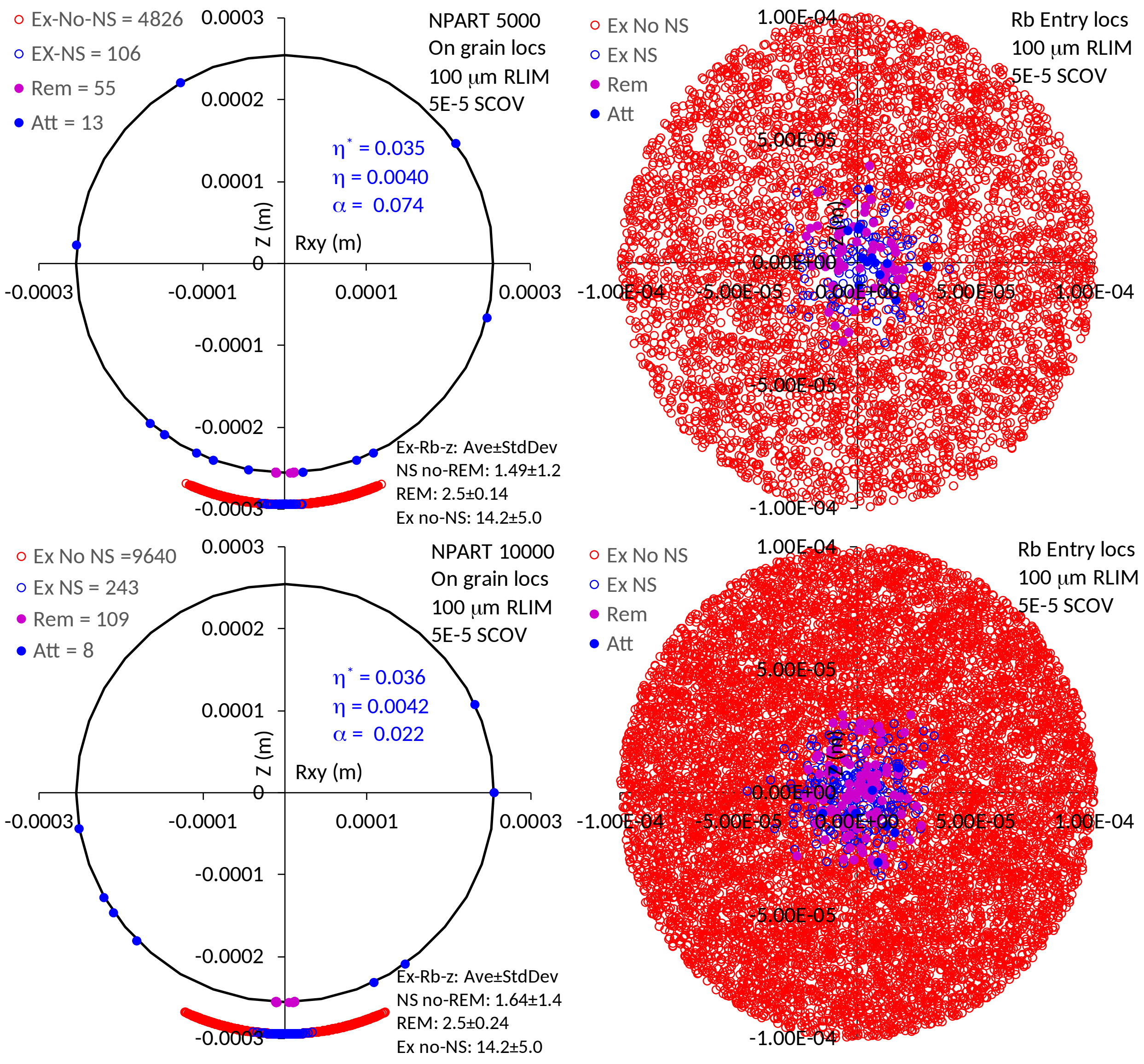

**Figure SI-10**. Traj-Hap outcomes among attachment, remained without attachment, exited the fluid shell with and without visiting the near surface domain (left side) and fluid envelope x-y entry locations (right side) comparing repeat simulation with 5000 simulated colloids (top row) and 10,000 simulated colloids (bottom row) with the same input parameters used in Traj-Hap simulations in the main text.

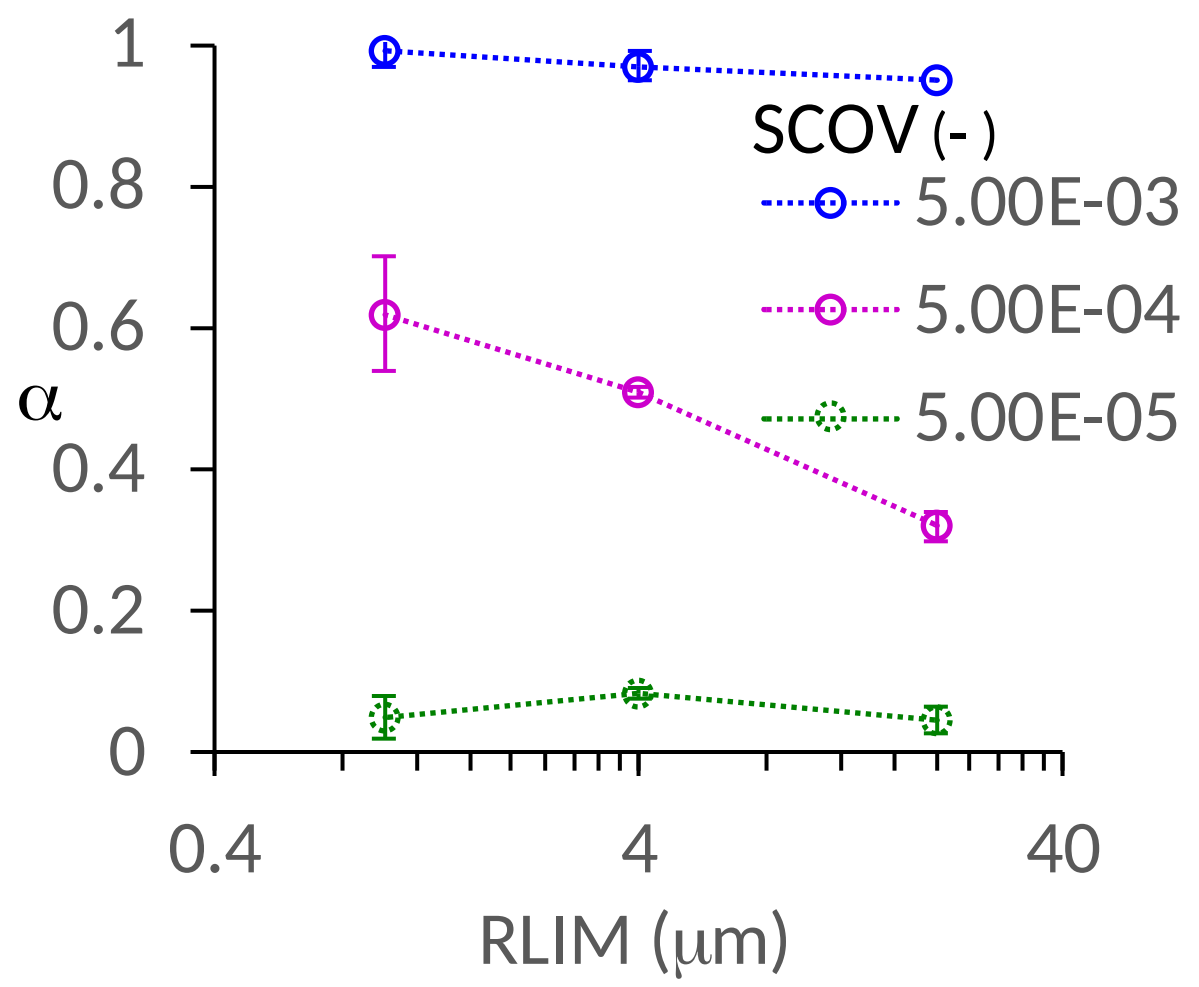

**Figure SI-10**. Values of $\alpha$ as a function of limiting radius of injection (RLIM) and surface coverage by heterodomains (SCOV) as determined in mechanistic trajectory simulations in the Happel sphere-in-cell collector.